# Interaction of Electromagnetic Radiation with Cometary Dust

## Ludmilla Kolokolova and Michael S. P. Kelley
### University of Maryland

## Hiroshi Kimura
### Planetary Exploration Research Center, Chiba Institute of Technology

## Thiem Hoang
### Korea Astronomy and Space Science Institute

The chapter overviews the recent developments in the remote sensing of cometary dust using visible, near-infrared, and thermal-infrared radiation, as well as interaction of the dust with electromagnetic radiation, which affects the dynamics of dust particles. It considers photometric, polarimetric, and spectral studies of cometary dust, focusing on those observables and correlations between them that allow revealing the composition, size, and structure of the dust particles. The analysis includes the observed brightness and polarization phase curves, color and polarimetric color of the cometary dust, and near- and thermal-infrared spectra. Special attention is paid to the role of gas contamination in the polarimetric and photometric data. A review of modeling efforts to interpret the observational results is also provided, describing the most popular (and some novel) techniques used in the computer modeling of light scattering by dust particles with a focus on modeling the most complex type of cometary particles: fluffy and porous agglomerates. The chapter also considers how properties of the dust particles affect their photoelectric emission and their response to the radiation pressure and radiative torque, including alignment and fragmentation of particles. Results of computer and some laboratory modeling are analyzed for their consistency with the observational and *in situ* data. Also discussed is how the modeling results can be combined with *in situ* data for better characterization of the cometary dust.

## 1. INTRODUCTION

Despite the great success of space missions studying comets, remote sensing using Earth-based observations (i.e., observations with groundbased and spacebased telescopes) remains the major source of information about comets. It allows studying a variety of comets, enabling further statistical analysis of their properties. Also, Earth-based observations allow monitoring the comets at different orbital positions and apparitions, thus providing information about both short-term and long-term evolution of the cometary environment.

Most of the cometary remote-sensing information is coming from observations of the interaction of cometary material with solar radiation. This chapter focuses on characterization of the dust in cometary coma based on how it scatters, absorbs, and emits radiation. Trying to avoid repetition of the facts and concepts considered in the previous volume, *Comets II* (*Kolokolova et al.,* 2004a, hereafter *K_C2*), this chapter emphasizes the new developments in the field. The observations and their results are overviewed in section 2. The chapter also summarizes the most efficient and broadly used techniques for interpretation of the observations (section 3). It demonstrates how characteristics of the cometary dust and their variations in the coma and along the orbit can be revealed using computational and laboratory modeling. Section 4 considers how the interaction with solar radiation affects motion, spinning, and fragmentation of the dust particles.

The main remote-sensing techniques that provide information about cometary dust are photometry and polarimetry in the visible and near-infrared (NIR), and spectroscopy in the NIR and mid-infrared (MIR) wavelengths. Most comets show similar photometric and polarimetric properties that allowed establishing the following regularities:

In the visible:

1. The color of the cometary dust is mainly red, with the values very similar for long-period and short-period comets and decreasing with increasing wavelength. Statistically averaged cometary color does not show stable tendencies with phase angle or heliocentric distance (see more in section 2.1).

2. Polarization of cometary dust is characterized by a negative (polarization plane is parallel to the scattering plane) parabolic-shaped branch at phase angles smaller than 20° and a positive (polarization plane is perpendicular to the scattering plane) bell-shaped branch at larger phase angles, reaching the minimum values at about –2% at the phase angle ~10° and the maximum values varying within 10–35% at phase angle ~90°–95°. For details, see section 2.2.

3. In the majority of comets, polarization increases with wavelength, although there are some exceptions.

4. Some comets demonstrate circular polarization. which usually does not exceed 2%.

5. The spectra of cometary dust show no absorption features.

In the NIR:

1. The increase of brightness with wavelength, i.e., red color, is also typical for the NIR; however, it is less red than in the visible

2. The dependence of polarization on phase angle in the NIR looks similar to the one in the visible for comets with high polarization maximum.

3. The majority of comets show a decrease of polarization with wavelength in the NIR.

4. Comets may demonstrate absorption bands of ice at 1.5, 2.2, and 3.1 μm, especially at large heliocentric distances.

In the MIR (thermal emission):

1. The main features in cometary spectra are silicate bands at 10 and 20 μm whose strength varies by comet and can vary with time.

2. The effective continuum temperature (based on the spectral shape) tends to be warmer (up to 30%) than the isothermal temperature of a large blackbody sphere at the same heliocentric distance.

The majority of the above regularities were obtained using aperture observations and were discussed in *K_C2*. The main advantage of the recent observations is that the aperture observations were replaced by imaging observations with a high spatial resolution. This brought an understanding that some of the listed regularities describe the cometary dust on average, whereas the dust experiences significant changes as the particles move out of the nucleus. Also, the dust was found to show different observational characteristics in cometary jets and other morphological features.

Another recent development in cometary dust studies is a huge progress in computational modeling of light scattering by non-spherical particles, specifically development of powerful computer packages to compute aggregated/agglomerated particles (see section 3). An important role in this development is played by an availability of more powerful computers, especially computer clusters, which allow reaching a significant increase in the efficiency of the computations by using parallelized computer codes, described in section 3.1. Very beneficial for the interpretation of Earth-based observations have been the results of laboratory and *in situ* studies of cometary dust, specifically based on the results of the Stardust and Rosetta missions. Information from those studies, reviewed in the chapters in this volume by Poch et al. and Engrand et al., allows preselecting some characteristics of the dust particles, specifically their size, structure, and composition. This narrows down the parameter pool involved in the modeling, thus increasing the modeling efficiency.

## 2. OBSERVATIONAL RESULTS

Interaction of cometary dust particles with solar radiation results in radiation being scattered, absorbed, and reemitted by the dust particles. This interaction varies depending on the location in the coma and distance to the Sun, but being averaged, produces the fundamental observational characteristics of the cometary dust: dependence of the brightness and polarization on phase angle (i.e., the Sun-comet-observer angle) and on wavelength.

In general, the measured brightness of the scattered light does not characterize the physical properties of the dust particles as it represents a combined effect of the intrinsic physical properties of the dust particles (size, composition) and their number, i.e., the column density of the dust particles. To cancel the effect of the column density and thus find out the intrinsic properties of the cometary dust, the observational characteristics defined by the ratio of the brightness — color and degree of linear polarization — are used. There are two other fundamental characteristics of the scattered light that are defined by the intrinsic dust properties: the dust albedo and its phase function.

In the light-scattering theory, the particle albedo is usually defined as a single-scattering albedo, i.e., the ratio of the particle scattering efficiency to its extinction efficiency (*van de Hulst*, 1957). However, this characteristic cannot be directly obtained from the observations. In cometary physics, albedo of a dust particle means its reflectivity and is analogous to the geometric albedo in planetary physics, i.e., it is the ratio of the energy scattered at phase angle equal to 0° to that scattered by a white Lambert disk of the same geometric cross section. The detailed description of how to determine the cometary dust albedo from observations as well as some other definitions of the particle albedo can be found in *K_C2*, section 2.1. The geometric albedo of a single particle is defined as $a_0 = \pi S/(Gk^2)$ (*Hanner*, 1981; *K_C2*) where S is the intensity of the light scattered by the particle in the direction of the Sun (zero phase angle) divided by the intensity of the solar light incident to the particle, G is the geometric cross-section of the particle, and k is the wave number related to the radiation wavelength λ as k = 2π/λ. It is clear from the definition that the geometric albedo is an intrinsic particle characteristic defined mainly by the particle composition and probably structure, as the intensity of the scattered light can be reduced by the scattering on the inhomogeneities within the particle. The parameter S depends on size, which in light scattering is characterized through the size parameter x = kr, where r is the particle radius, but for large particles S/G becomes constant, which makes the effect of size negligible for particles of x > 10 (*Mishchenko and Travis,* 1994).

Often the albedo of a particle at other than zero phase angles is considered, i.e., $a(\alpha) = \pi S(\alpha)/(Gk^2)$, where $\alpha$ is phase angle. This is also an intrinsic property of the particle, and $a(\alpha)/a_0$ is identical to the phase function, which demonstrates how the reflectivity of the particle changes with phase angle. Often the phase curve is defined as the normalized intensity curve, i.e., the observed brightness at each phase is divided by the brightness at zero phase angle. In this case the phase curve is identical to $a(\alpha)/a_0$, thus, $a(\alpha)$ is the particle geometric albedo multiplied by the phase function.

The phase curve of the particles strongly depends on their size parameter (*van de Hulst,* 1957; *Bohren and Huffman,* 2008). Small particles with x < 1 scatter light in the Rayleigh regime; their phase function is symmetric, smoothly decreasing from 0° to 90° and then smoothly increasing. For larger particles, the curve becomes more asymmetric with forward scattering increasing with size. Monodisperse compact particles with x between 1 and 20 usually have a phase curve with a complex oscillating structure, which disappears for larger particles where light scattering occurs in the geometric-optics regime. The large particles usually have a strong forward-scattering peak; at smaller phase angles, the curve becomes less steep and at some sizes and composition may show a backscattering peak. The current envision of the cometary phase curve mainly attributes to the composite phase curve compiled by *Schleicher* (2010) based on an empirical function fit to Comet 1P/Halley data from *Schleicher et al.* (1998) and a Henyey-Greenstein model fit to near-Sun comet photometry by *Marcus* (2007). It has a medium-backscattering peak after which the brightness decreases, reaches minimum at about 50°–60° and then increases, becoming steeper as it approaches the forward scattering. Similar phase curves were found for zodiacal dust, Saturn's rings, and protoplanetary and debris disks (Fig. 1) (see also *Hughes et al.,* 2018). A smooth shape with a large forward-scattering and some backscattering peaks can be attributed to a polydisperse distribution of the particles comparable with wavelength, i.e., of size parameter x = 1–5 (*K_C2*).

A rather different shape of the phase curve for Comet 67P (hereafter 67P) with the minimum around 100°, a steeper increase in the backscattering direction, and a rather gradual increase in the direction to the forward scattering (*Bertini et al.,* 2017) was acquired during the Rosetta rendezvous mission. *Moreno et al.* (2018) explained this unusual curve by light scattering on oriented elongated particles. This explanation was supported by a specific position of the camera that was observing the particles affected by the gas flow and radiation torque (see section 4.3). However, later a similar phase curve was observed for some debris disks (*Hughes et al.,* 2018) (see Fig. 1 for examples). The laboratory measurements by *Muñoz et al.* (2020) showed that such curves are typical for large, millimeter-sized, porous particles that can be a reasonable explanation for the case of near-nucleus particles in Comet 67P where the near-nucleus coma is dominated by particles hundreds of micrometers in size (see the chapter in this volume by Engrand et al.).

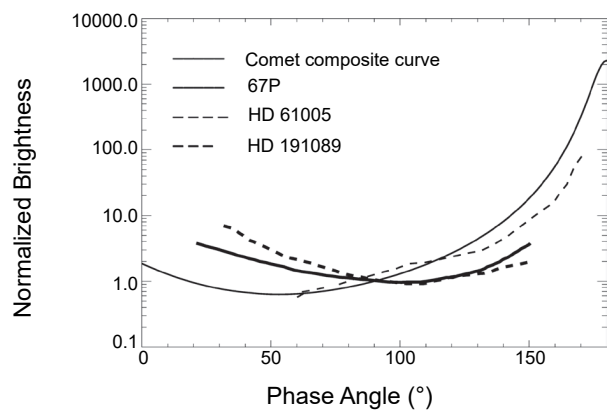

**Fig. 1.** Phase curves for comets and debris disks. A thin solid line shows the composite dust phase function for comets by *Schleicher* (2010). The thick solid line is the phase curve for Comet 67P measured by the Rosetta OSIRIS camera (*Bertini et al.,* 2017). Dashed lines show the phase curves for debris disks HD 61005 [thin line (*Olofsson et al.,* 2016)] and HD 191089 [thick line (*Ren et al.,* 2019)]. Two types of the phase curves can be seen for both comets and debris disks.

## 2.1. Color of Cometary Dust

The color of cometary dust is traditionally defined as the difference of the magnitudes in two continuum filters, e.g., $m_c = m_B - m_R$ in the case of B-band and R-band filters. Since the magnitude is –2.5 log I where I is the brightness in physical units, the color can be presented as –2.5 log($I_B/I_R$). As it is a ratio of the brightness in two filters, it cancels the effect of dust particle column density and becomes a characteristic of the dust particles per se. It is worth noting that the presented definition of the color contains information on the solar spectrum as it represents the solar light scattered by the comet. Subtraction of the solar color calculated for the same filters, i.e., $m_s = m_{sB} - m_{sR}$, from the observed dust color, gives the intrinsic color of the cometary dust equal to $m_c - m_s$.

The other popular definition of the cometary dust color uses the gradient of the cometary spectrum, i.e., the steepness of the cometary continuum. Presenting the cometary spectrum as $S(\lambda) = F_c(\lambda)/F_{Sun}(\lambda)$, i.e., the comet spectrum $F_c(\lambda)$ divided by the solar spectrum $F_{Sun}(\lambda)$, one can calculate cometary color (*A'Hearn et al.,* 1984) as the normalized gradient of reflectivity:

$$S'(\lambda_1,\lambda_2) = \frac{2000}{\lambda_2 - \lambda_1} * \frac{S(\lambda_2) - S(\lambda_1)}{S(\lambda_2) + S(\lambda_1)} \quad (1)$$

where $S(\lambda_1)$ and $S(\lambda_2)$ correspond to the dust reflectivity at the wavelengths $\lambda_1$ and $\lambda_2$ under the condition $\lambda_2 > \lambda_1$. $S'(\lambda_1, \lambda_2)$ is expressed in percent per 1000 Å.

Progress in studying cometary colors is characterized by two main developments: (1) collection of many new values of the colors for different comets and at their different positions on the orbit that allow accomplishment of a statistical analysis of the cometary colors; and (2) acquiring images of the brightness in different filters, thus producing maps of cometary colors that reveal changes in the color with the distance from the nucleus and in different cometary features. The color variations in the coma will be considered

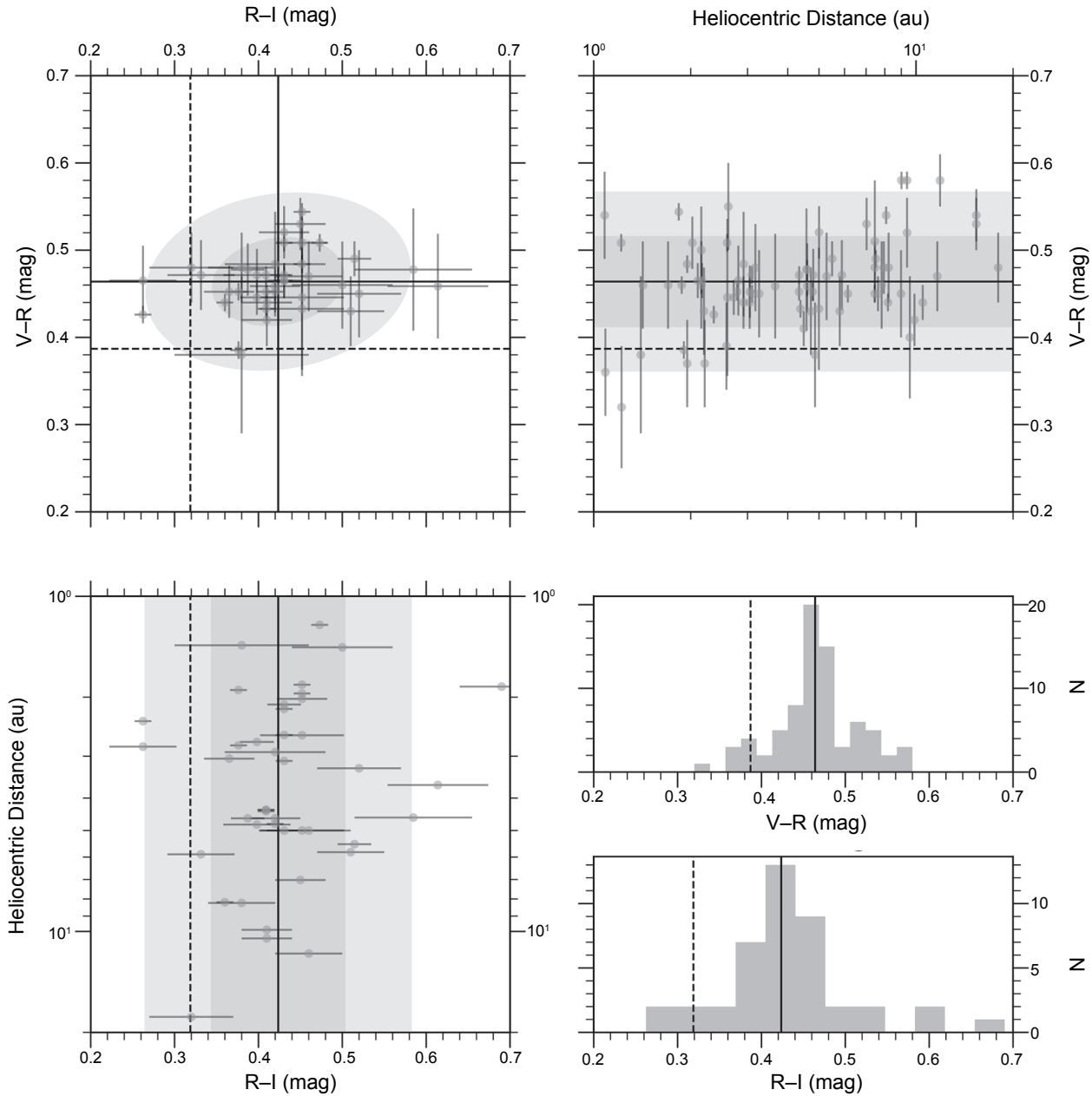

**Fig. 2.** Broadband coma colors based on photometry in the V, R, and I bandpasses, or transformed to V, R, I from equivalent filters. Solid lines are the mean colors, dashed lines are solar colors. Shaded regions indicate 1 and 2 standard deviations from the mean colors. *Upper left:* Color-color plot and covariance ellipses. *Upper right and lower left:* All V–R and R–I data points vs. heliocentric distance. The axes are configured so that individual points in the V–R vs. R–I plot may be traced to the color-heliocentric distance plots. *Lower right:* Color histograms.

in section 2.2.1.

Observations of cometary coma continua span a moderate range of colors, from slightly blue to red. We searched the literature for surveys of broadband optical colors of active comets. Based on the results of *Solontoi et al.* (2012), we added the requirement that any color uncertainties must be better than 0.1 mag. In Fig. 2, we present V–R and R–I colors from *Solontoi et al.* (2012), *Jewitt* (2015), and *Betzler et al.* (2017). A minimum uncertainty of 0.05 mag was adopted for all Beltzer et al. data based on Anderson-Darling tests, which could not distinguish between true variability and a normal distribution for the six comets with time-resolved observations.

Cometary colors are typically measured with broadband filters, which can be affected by the presence of gas emission bands. For example, using spectra of the dusty Comet C/2013 $US_{10}$ (Catalina), *Kwon et al.* (2017) estimated a moderate gas contamination of 8% and 3% in the R- and I-bands ($9500 \times 1900$ km aperture). Observations of gas-rich Comet 252P/LINEAR in the V, r', BC, and RC filters by *Li et al.* (2017) imply a photometric excess of 25% to 50% in the V and r' filters due to the presence of gas (2300-km aperture radius, rh = 1.1 au). Therefore, narrowband filter sets designed for cometary comae are better suited for continuum colors; the most recent iteration is the HB filter set (*Farnham et al.,* 2000). But even the narrowband filters may be significantly contaminated. For the HB filter set, $C_3$ and OH emission is found within the UC filter bandpass ($345 \pm 40$ nm), and $C_2$ in the GC filter (*Farnham et al.,* 2000; *Rosenbush et al.,* 2002; *Opitom et al.,* 2015).

Gas contamination will be stronger for comets with lower dust-to-gas production rates. *A'Hearn et al.* (1995) presented the ratio of Afρ (the product of albedo, aperture filling factor, and aperture radius, an empirical proxy for dust production rate) and OH production and found that comets span the range $\log_{10}(Af\rho/OH) = -26.5$ to $-24.9$ for units of cm s molecule$^{-1}$ (note that these values are not corrected for dust phase angle effects), thus comets on the low end of the range are said to be gassy and those on the higher end are dusty. Most optical gas emission bands are from molecules produced by photolysis in the coma, which is a function of heliocentric distance. The emissions from these bands have spatial profiles that tend to be shallower than dust spatial profiles (*Combi et al.,* 2004), thus the physical size of the aperture at the distance of the comet also plays an important role.

Altogether, gas contamination is a general problem in photometry of cometary dust, and the cometary astronomer should consider the possibility of gas emission in their data. This statement is especially true when observing unusual comets or comets in unusual circumstances, e.g., see *Bellm et al.* (2019) and *McKay et al.* (2019) for CO$^+$ emission in g-band and HB UC filter images of Comet C/2016 R2 (PanSTARRS). Of all the optical broadband filters, the redder filters ($\lambda > 600$ nm) tend to be the least affected by gas, but $NH_2$ and CN bands may still be significant (*Fink,* 2009). Gas contamination can also be estimated or avoided with spectroscopic measurements (e.g., *Jewitt and Meech,* 1986).

The average and standard deviation broadband colors are V–R = $0.46 \pm 0.05$ mag (N = 66) and R–I = $0.43 \pm 0.08$ mag (N = 40). Figure 2 shows histograms of the V–R and R–I color sets. The distributions are unimodal and nearly symmetric.

Coma broadband colors may be expected to have a heliocentric distance dependence due to gas contamination as discussed above, the presence of water ice, or the variation of the coma grain size distribution. We compared data taken inside and outside of different heliocentric distance cuts, ranging from 1.5 to 10 au, with the two-sided Kolmogorov-Smirnov test. After accounting for uncertainties, all p-values were >5%, indicating none of the tests found significant differences, consistent with prior results (*Solontoi et al.,* 2012; *Jewitt,* 2015). This dataset may not be useful for realizing these effects if they are much smaller than the observed scatter in the population (~0.08 mag). Moreover, gas contamination may not be apparent given that so few data points are taken near 1 au. Investigating these effects likely requires isolated studies of individual comets with precise color measurements.

In contrast with the groundbased data, spacecraft observations of Comets 67P and 103P/Hartley 2 have shown clear color trends with distance to the nucleus and, for 67P, with distance to the Sun and phase angle. *Filacchione et al.* (2020) show that the optical (0.5 to 0.8 μm) spectroscopic color of the inner (1 to 2.5 km) coma of 67P varied with heliocentric distance, from near 20% to 25%/0.1 μm at 1.2 au, to near 0% to 5%/0.1 μm at 2.8 au, consistent with the presence of icy grains in the coma. This magnitude of a change should be easily observed in groundbased data, but the analysis of *Kwon et al.* (2022) suggests that this variation is limited to the innermost coma of this comet. *Bockelée-Morvan et al.* (2019) show a phase angle dependence of Comet 67P's inner coma at 2 to 2.5 μm. The spectral gradient's slope, measured with respect to the continuum flux density at 2 to 2.5 μm, was 0.031%/100 nm/° from 50° to 120° and thus demonstrated a noticeable reddening. They point out that that phase reddening is common on solar system bodies and has also been observed in the zodiacal light, and argue that phase reddening in a cometary coma is likely caused by porous particles much larger than the wavelength of light being observed (i.e., >10 μm in size). Phase reddening for dust particles was indicated in the laboratory measurements by *Escobar-Cerezo et al.* (2017) and was also attributed to the particle structure (roughness) by analogy with the explanation of the phase reddening for rough surfaces (*Schröder et al.,* 2014).

Further examination of the potential effect of water ice on coma optical colors became possible due to the Deep Impact observations of Comet 103P/Hartley 2. Near-infrared observations of this comet showed prominent water ice absorption bands at distances within a few kilometers from the nucleus. *Protopapa et al.* (2014) examined these absorption features in an ice-rich spectrum taken 400 m from the surface and estimated an ice abundance of 5% by area. *La*

*Forgia et al.* (2017) studied the optical colors of this comet and showed that the ice-enriched regions tend to be bluer than the ice-poor regions in the coma. The azimuthally averaged spectral gradient from the green to red (0.53 to 0.75 μm) ranges from 9% per 0.1 μm near the nucleus to 13% per 0.1 μm 40 km from the nucleus. Given that the water ice is asymmetrically distributed in the coma, the color difference would be even larger if it were not azimuthally averaged. The quoted range corresponds to V–R colors of 0.49 to 0.54 mag, which is fully on the red end of the observed coma colors (Fig. 2), emphasizing that ice is difficult to identify based on absolute colors alone. However, the 0.05-mag change in color is large enough to be apparent in the observed V–R colors of cometary comae. The lack of strong bluing with heliocentric distance in Fig. 2 may suggest that ice is rare in cometary comae, but considerations for competing effects are necessary. The sublimation lifetime of ice compared to the photometric aperture size is of foremost importance. There is some evidence for color gradients in distant comae, reddening by 3% to 8%/0.1 μm with distance from the nucleus, which is suggestive of the presence of water ice grains (*Fitzsimmons et al.*, 1996; *Li et al.*, 2013, 2014).

A changing grain size distribution may also affect coma dust color. As activity increases and overcomes the tensile strengths of the dust agglomerates (*Gundlach et al.*, 2015), small grains could populate the coma, producing a bluer color. Alternatively, grain fragmentation may increase the number of Rayleigh scatterers with increasing nucleus distance.

There are several papers reporting narrowband photometry of cometary comae. We searched the literature for observations with the HB filter set and assembled data on 19 comets from 18 papers (*Bair et al.*, 2018; *Cudnik*, 2005; *Farnham and Schleicher*, 2005; *Ivanova et al.*, 2014; *Knight and Schleicher*, 2013; *Knight et al.*, 2021; *Li et al.*, 2017; *Moulane et al.*, 2018, 2020; *Opitom et al.*, 2015; *Zhang et al.*, 2021; *Schleicher*, 2007, 2008, 2009; *Schleicher and Bair*, 2011; *Schleicher et al.*, 2003). We chose the HB filter set, given its improvements over previous narrowband filter sets, especially with respect to gas contamination mitigation. Figure 3 presents all colors with uncertainties <0.1 mag in color-color plots as a linear spectral slope (equation (1)). It shows that the observations span a moderate range of spectral slopes, from neutral to moderately red (0% to 30%/0.1 μm). However, Comet 96P is a clear outlier with a UC–BC slope of –18%/0.1 μm in accordance with the findings of *Opitom et al.* (2015), that the OH (0–1) band may be a significant contributor to UC filter photometry. A crude correction to the UC reflectance using the relative strength of the OH (0–1) and (0–0) bands (*Schleicher and A'Hearn*, 1988) increased the comet's mean spectral slope from –18 to –4%/0.1 μm. A near-UV spectrum of 96P may provide some clarity on the UC–BC colors of cometary comae, as would further investigations into the magnitude of OH contamination in the UC filter.

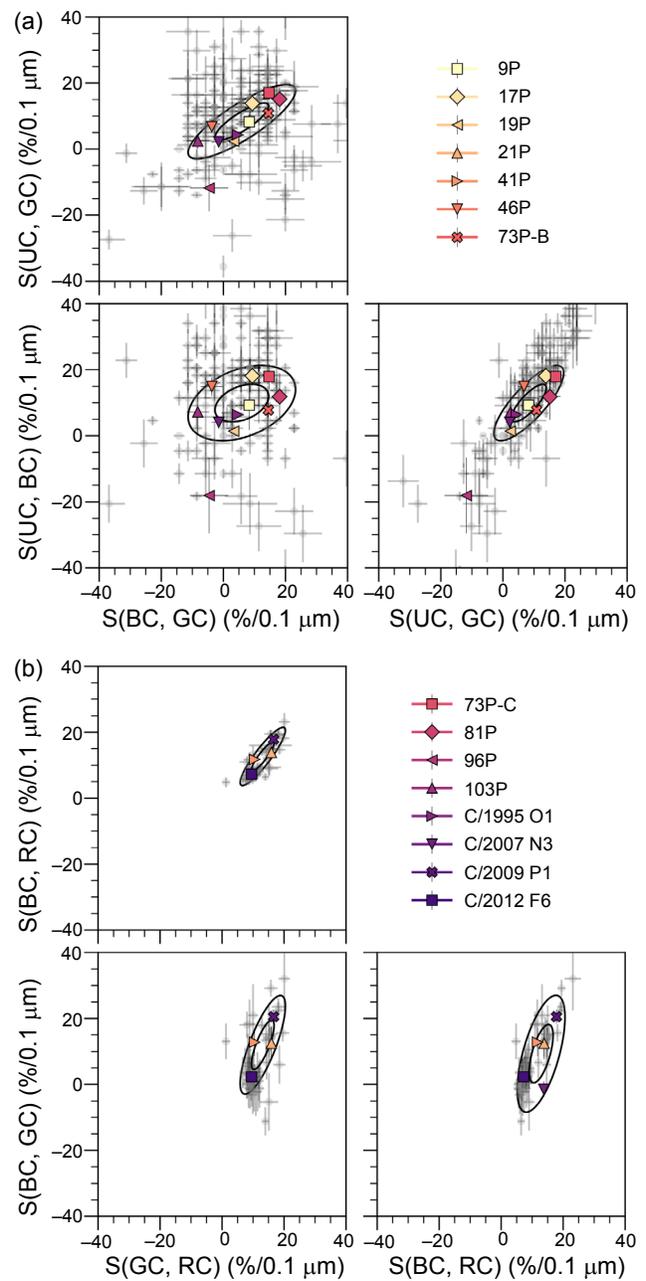

**Fig. 3.** Narrowband colors of comets based on the HB filter set. Individual measurements with uncertainties better than 10%/0.1 μm are shown as light gray circles. Comet-by-comet averages are also plotted; uncertainties are based on the error on the mean, or the standard deviation of the data, whichever is greater.

### 2.2. Cometary Polarimetry

In planetary astronomy, the degree of linear polarization (P, hereafter called polarization) is defined as

$$P = (I_\perp - I_\parallel)/(I_\perp + I_\parallel) \qquad (2)$$

where $I_\perp$ and $I_\parallel$ are the intensity components perpendicular and parallel to the scattering plane (the plane that contains the Sun, comet, and observer). Based on this definition, the polarization with $I_\perp > I_\parallel$ is called positive, and if $I_\perp < I_\parallel$ it is called negative. This definition is given here mainly to explain the terms negative and positive polarization and make it clear that polarization is the ratio of intensities and thus depends only on the properties of the dust particles. Currently, with the rise of the Stokes polarimetry (e.g., *Keller et al.,* 2015), polarization is usually defined through the Stokes vector, a mathematical object that is formed by components (I, Q, U, V), where I is the total intensity of the scattered light, Q is the numerator from equation (2), and Q/I is identical to the planetary definition of the linear polarization; U together with Q defines the angle θ that determines the position of the polarization plane (i.e., the plane in which oscillates the linearly polarized part of the electromagnetic wave) as tan(2θ) = Q/U. Note that a physically correct way to define the linear polarization is through $P = (Q^2 + U^2)^{1/2}/I$ and angle θ; however, in the planetary observations a typical value of θ is close to 0 (negative polarization) or 90° (positive polarization), which justifies the definition given in the beginning of the section. Parameter V defines circular polarization, i.e., the part of the scattered light for which the plane of the electromagnetic wave rotates around the direction of the light propagation. More details on the Stokes vector can be found in *Bohren and Huffman* (2008) or in *K_C2*. The value of the Stokes vector representation is that it allows using matrix algebra to consider light scattering by complex systems. Specifically, the Stokes vector of the radiation source $(I_0, Q_0, U_0, V_0)$ is related to the measured Stokes vector through the 4 × 4 Mueller matrix (often called scattering matrix) as (I, Q, U, V) = **M**$*(I_0, Q_0, U_0, V_0)$; for the solar radiation the Stokes vector is $(I_0, Q_0, U_0, V_0) = (I, 0, 0, 0)$.

A detailed review of the cometary polarimetric studies can be found in *Kiselev et al.* (2015); below we briefly outline the results reported there, adding the results published after that review.

A comprehensive collection of the results of the cometary polarimetric observations has been recently archived by the NASA Planetary Data System (PDS) as the Database of Cometary Polarimetry (*Kiselev et al.,* 2017). It includes published and unpublished data of 3441 observations for the period 1881–2016. The database includes data for 95 comets with the description of the geometry of observations and characteristics of the polarimetric instrumentation. The data cover the wavelength from 0.26 to 2.32 μm and the ranges of phase angles, heliocentric distances, and geocentric distances 0.0–122.1°, 0.0–7.01 au, and 0.01–6.52 au, respectively. The data from this database and its earlier version (*Kiselev et al.,* 2006) have been already used in several studies where some statistical analysis of the polarimetric data was done; see, e.g., *Mishchenko et al.* (2010), *Borisov et al.* (2015), *Kwon et al.* (2019, 2021), *Rosenbush et al.* (2021). The most interesting observations acquired since the release of the database will be discussed in this section. They have not changed the main characteristics of the dependence of polarization on wavelength and phase angle with the probable exception of the data for interstellar Comet 2I/Borisov (*Bagnulo et al.,* 2021), which showed the values of polarization exceeding the values typical for solar system comets, although similar to those observed for Comet C/1995 O1 Hale-Bopp.

Probably the most significant change in our understanding of the cometary polarization since *K_C2* is reconsideration of the classification of comets in two polarimetric classes: comets with a high polarization maximum (~25%) and those with a low polarization maximum (~10%) (*Chernova et al.,* 1993; *Levasseur-Regourd et al.,* 1996). *Chernova et al.* (1993) related these two classes to different dust/gas ratios (see definition in section 2.1) in the comets; specifically, high-polarization comets were found to have a large dust/gas ratio, and that for the low-polarization comets was at least twice lower. This allowed relating the polarization difference in those two classes to a stronger depolarizing effect of the gas emissions in the low-polarization comets. With many more data acquired after 1996, one can see (Fig. 4) that there is no clear division of the comets into two classes. Although they may be still visible in the blue domain, the comets observed in the red and NIR domain tend to form a single curve. There is a scattering of the data, especially in the blue domain. The main explanation of the scattering relates to the original division of the comets to dusty and gassy types. However, when narrowband filters weakly affected by the gas emissions are used, all comets show similar polarization, typical for the dusty comets.

One more confirmation of the significant influence of gas contamination on the polarimetric data was obtained from observations of cometary polarization with increasing aperture of the observations (*Jockers et al.,* 2005) or using imaging polarimetry (*Kwon et al.,* 2017). These observations demonstrated that for both types of comets the near-nucleus polarization is high, reaching the values 20–25%. However, with increasing aperture or distance from the nucleus, the values of polarization in dusty comets change only slightly, whereas for the gassy comets, a decrease in polarization is significant, bringing the polarization to the values typical for low-polarization comets or even lower, as shown in Fig. 5. *Jockers et al.* (2005) concluded that the low values of polarization of gassy comets resulted from the contamination of the data by gas emissions, which depolarize the light.

Recent observations and detailed analysis of the gas contamination have been reported by *Ivanova et al.* (2017), *Kwon et al.* (2017, 2018), and *Kiselev et al.* (2013, 2020). Spectropolarimetric data for Comet Garradd (*Kiselev et al.,* 2013) showed correlation between the intensity of the gas emissions and polarization. Similarly, *Kwon et al.* (2017) found a strong correlation between the polarization and the intensity of gas emissions in their change with the distance from the nucleus. Obviously, the effect of gas contamination should increase with the distance from the nucleus as the amount of the gas from the distributed gas sources adds to the original gas sublimated from the nucleus.

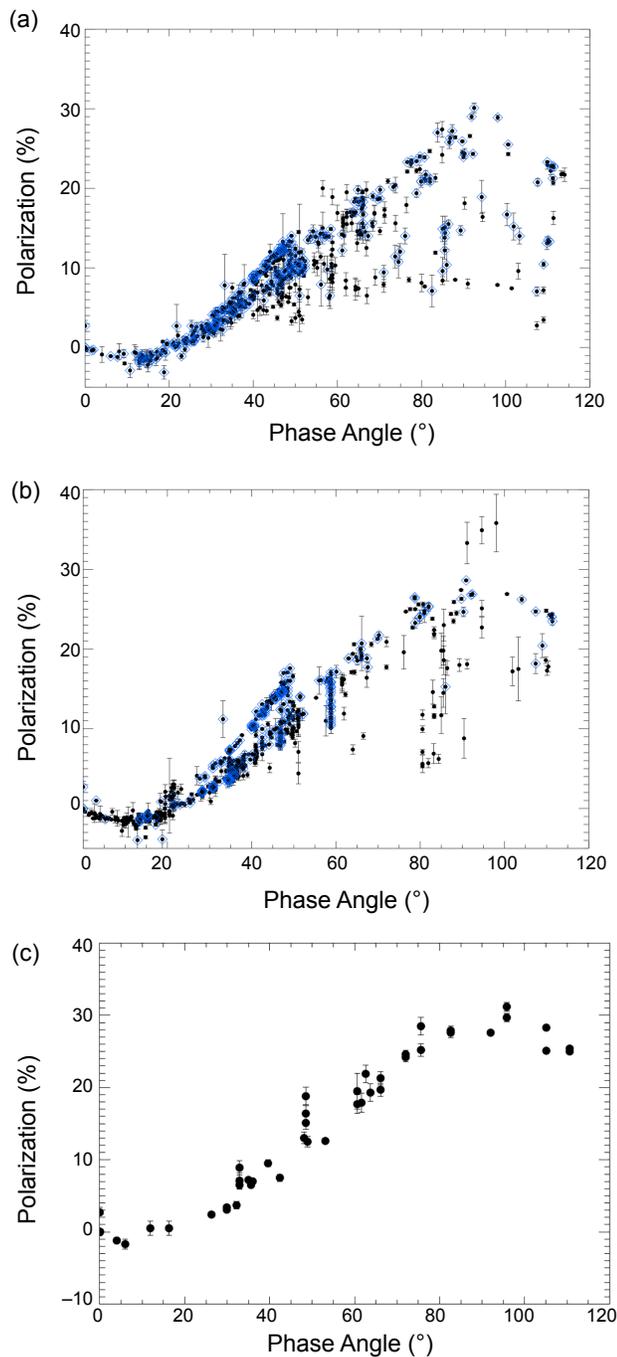

**Fig. 4.** Compilation of the data from *Kiselev et al.* (2017) for the cometary polarimetric phase curve. **(a)** Data acquired in the blue domain (blue squares are for the BC_IHW filter). **(b)** Data in the red domain (blue squares are for RC_IHW filter). **(c)** Data in the NIR region (JHK filters); the data for Comet West are excluded as they were strongly affected by the thermal emission (*Oishi et al.*, 1978).

The dependence of the polarization on the aperture of observations, which can move a comet from the high-polarimetric class to the low-polarimetric class, makes the division of the comets into two classes questionable, or at least not reflective of the intrinsic properties of the cometary dust. However, a more detailed analysis and especially correlation of the positive polarization of comets with the strength of the silicate infrared feature (see section 2.3.2) allowed the assumption (*Kolokolova et al.,* 2007) that the difference in polarization may reflect a different distribution of the dust particles in the coma; specifically, in the low-polarization comets, the dust is concentrated near the nucleus, thus resulting in a low dust/gas ratio for the overall coma. Such a difference in the dust distribution may result from the different size and porosity of the dust particles, as is shown in section 2.3.2.

An even more detailed picture of the change in polarization throughout the coma came from charge-coupled device (CCD) images with high spatial resolution. Early polarimetric images were reported by *Renard et al.* (1996), then a detailed study of Comet Hale-Bopp with the imaging polarimetry was published by *Jockers et al.* (1997) and *Hadamcik et al.* (1997). A paper by *Hadamcik and Levasseur-Regourd* (2003) presented results of imaging polarimetry for nine comets and showed that an area of lower positive polarization was observed in the near-nucleus area; it was named a "polarimetric halo." A polarimetric halo was observed at small phase angles as an area of more negative polarization in Comet ISON (*Hines et al.,* 2013) and other comets (*Choudhury et al.,* 2015). Since the polarimetric halo has only been observed in some comets, its origin is still not clear. The majority of high-resolution polarimetric images of comets show a rather gradual change in polarization. It appears that for some comets polarization decreases with the distance from the nucleus, probably indicating increasing gas contamination; however, for some other comets polarization increases. Increase in the positive polarization is likely a manifestation of dust particle fragmentation (e.g., *Jones et al.,* 2008) producing an increased abundance of small particles that scatter light in the Rayleigh regime. Both gas emissions and Rayleigh scatterers produce a bell-shaped phase curve that has positive polarization at all phase angles (e.g., *Kiselev et al.,* 2015), thus at low phase angles they can cause a decrease in the magnitude of negative polarization. An increase in the absolute value of the negative polarization with the distance from the nucleus, if observed, has a more complicated origin. Since negative polarization is most likely a result of multiple scattering inside a fluffy/porous particle (e.g., *Muinonen et al.,* 2012), an increase in absolute value of the negative polarization may indicate increasing transparency of the particles due to either evaporation of some dark material (e.g., semivolatile organics) or increasing porosity (*Kimura et al.,* 2006). There are also more complicated cases. For example, Comets 2P/Encke (*Jewitt,* 2004) and 67P (*Rosenbush et al.,* 2017) demonstrated a decrease in polarization that changed to an increase at larger distances from the nucleus, as shown in Fig. 7. A possible explanation of such a behavior is provided in section 2.2.1.

For a detailed discussion of the polarization in the cometary jets and similar features, the reader can check the review by *Kiselev et al.* (2015). Although most often the polarization in the jets is higher than in the ambient coma, suggesting particles of different composition or size in the jets and in the ambient coma, high-resolution images reveal that not all jets seen in photometric images are visible in polarization — in this case the particles in the jets and in the ambient coma are similar and the difference between the jets and the coma is in the number density of particles. Examples of both cases were found in Comet C/2011 KP36 (Spacewatch) (*Ivanova et al.,* 2021).

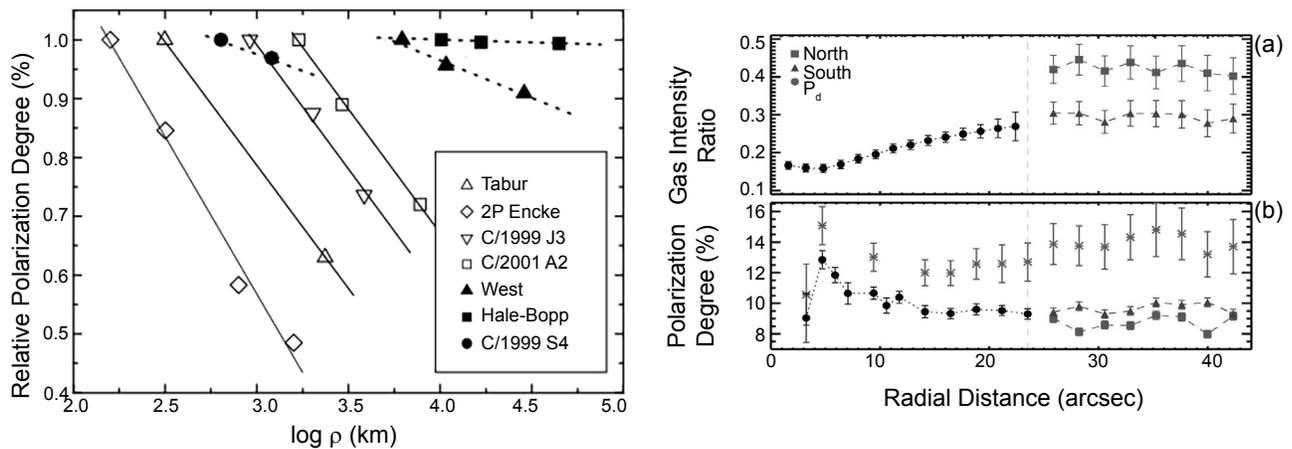

**Fig. 5.** Change of polarization with the distance from the nucleus. On the left, examples for low-polarization (white symbols) and high-polarization (black symbols) comets (*Kolokolova et al.,* 2007). On the right, the radial profiles of **(a)** the gas intensity ratio, f$g_{Rc}$, and **(b)** polarization for Comet C/2013 US$_{10}$ Catalina from *Kwon et al.* (2017); on the left of the gray dashed line at 23″ are aperture polarimetry results, and on the right of this line the sky regions were divided to the north and south wings and were treated separately. The gray asterisks in **(b)** denote polarization corrected with the corresponding values of **(a)**.

We also want to point a significance of the contribution of the cometary nucleus to the light in the near-nuclear coma. The contribution of the nucleus strongly affected the coma polarization and color up to a distance of 10,000 km from the nucleus for Comet C/2011 KP36 (*Ivanova et al.,* 2021) and up to 3000 km for Comet Encke (*Kiselev et al.,* 2020).

We cannot leave out the polarization measurements performed by the polarimeters onboard the STEREO spacecraft. *Thompson* (2015, 2020) reported that the polarization properties of Comets C/2012 S1 (ISON) and C/2011 W3 (Lovejoy) have a high positive polarization, reaching 60%, and a broad negative branch reaching out to an inversion point between 40° and 50°. Also, dramatic changes in polarization were observed as Comet ISON approached the Sun to distances closer than 10 solar radii, demonstrating that the dust in the Sun-grazing comets undergo significant changes in the vicinity of the Sun. To explain the STEREO data, *Thompson* (2020) suggests fragmentation and sublimation of olivine at those distances. This is in accordance with *Kimura et al.* (2002b) who showed that silicates, such as olivine and pyroxene, sublimate at heliocentric distances below 10 solar radii. The other mechanism can be the dust fragmentation as it is disrupted by the radiative torque (see section 4.3.3).

**2.2.1. Polarization of cometary dust in the near-infrared.** Adding the polarimetric data in the NIR has re-sulted in two important conclusions. First, the polarization phase curves in the NIR appeared to be almost identical to those in the visible (see Fig. 4). This indicated that the characteristics of the particles, which affect the polarization, did not change much as the observations move from the submicrometer to micrometer scale, most likely signaling that the dominating size parameter of the particles exceeds unity not only in the visible but also in the NIR, i.e., cometary dust particles are several micrometers or larger in size. The only exception is Comet Hale-Bopp, which showed almost absent negative polarization in the NIR. This behavior can be evidence for particles approaching the Rayleigh regime in the NIR, thus particles <1 μm in size.

The other special feature of the NIR polarization is that the spectral trend in polarization changes from positive to negative, i.e., polarimetric color, defined as P($\lambda_1$)–P($\lambda_2$) with $\lambda_1 > \lambda_2$, which is usually positive (red) in the visible (see *K_C2*), becomes negative (blue) in the NIR. Explanation of the spectral increase of polarization in the visible is quite straightforward: With increasing wavelength the dust particles or grains (monomers) in agglomerates are approaching the Rayleigh regime of scattering, and that manifests itself in increasing positive polarization. A spectral decrease in polarization requires a different explanation. *Kolokolova and Kimura* (2010a) noted that light scattering on a collection of closely located particles (e.g., monomers in agglomerates) can cause depolarization due to electromagnetic interaction between the monomers (cf. multiple scattering), which is stronger if the number of particles covered by a single wavelength is larger. Approaching the NIR, the wavelength is getting larger and thus it can cover more monomers, decreasing the polarization. This phenomenon was considered in detail in *Kolokolova et al.* (2011), where it was shown that it strongly depends on the porosity: The larger the porosity of the particle, the longer the wavelength at which the electromagnetic interaction becomes significant enough to cause depolarization of light. In a compact agglomerate the longer the wavelength the more monomers it covers, so the interaction between the monomers becomes stronger and the light becomes more depolarized. This results in a decrease of polarization with wavelength (blue polarimetric color). For a porous agglomerate, the number of monomers covered by a single wavelength does not change much as the wavelength increases, i.e., the change in the interaction between the monomers cannot overpower the change in the monomer size parameter, and the polarimetric color stays red. However, as the wavelength reaches some critical value, the number of covered monomers in the porous agglomerate changes significantly and interaction becomes the main factor that defines the polarimetric color, which then becomes blue. Thus, the wavelength at which the decrease in the polarimetric color starts may be determined by the porosity of the agglomerate, and it is smaller for more compact particles. This provides a straightforward explanation of the blue polarimetric color observed in the visible for some comets (*Kiselev et al.,* 2008): Most likely, the dust in those comets is dominated by more compact particles.

**2.2.2. Circular polarization.** As mentioned above, polarization can also have a circular component. Circular polarization was observed in more than 10 comets and was found at the tenths of percent level (Fig. 6). In the majority of cases, the absolute value of circular polarization (lefthanded polarization is usually considered as negative, and righthanded as positive) increases with the distance from the nucleus (see details in *Kiselev et al.,* 2015). The causes of circular polarization should be within processes or properties of particles that produce a mirror asymmetry of the medium (*van de Hulst,* 1957). Several mechanisms were suggested to explain the origin of circular polarization, among them such an exotic mechanism as homochirality of the molecules in cometary organics. However, T-matrix modeling of light scattering by homochiral molecules (see section 3.1) showed the values of circular polarization several orders lower than the observed values even in the case of the particles consisting entirely of homochiral organics (*Nagdimunov et al.,* 2013; *Sparks et al.,* 2015). More recent studies attribute cometary circular polarization to the alignment of the dust particles under solar magnetic field and radiative torque (see section 4.3).

**2.2.3. Correlation between color and polarization.** An efficient approach in studying properties of cometary dust is combining polarimetric data with observations of the dust color. Color and polarization depend solely on the intrinsic properties of dust particles, specifically their size and composition. Polarization is a more complex characteristic as it also depends on shape and structure of particles; however, if observations show a correlation between color and polarization, this can provide information on the composition or size of the dust particles.

For example, an increase in polarization together with a decrease (bluing) of the color (anticorrelation) was observed for Comet Encke starting with the distances ~3 arcsec (*Jewitt,* 2004) and Comet 67P (*Rosenbush et al.,* 2017) (see Fig. 7). These variations can be explained by a decreasing size of the dust particles (*Kolokolova et al.,* 2002). However, closer to the nucleus, both Encke and 67P showed a decrease in polarization accompanied by a decrease in color (correlation). The currently suggested explanation is that large particles (hundreds of micrometers) dominate close to the nucleus, then they fragment to smaller ones (~10 μm) that are still far from the Rayleigh particles; their color and polarization correlate. Then fragmentation continues, and as particles approach the Rayleigh regime, color and polarization anticorrelate. This scenario is confirmed by the laboratory simulations of light scattering by cometary analog particles (*Hadamcik and Levasseur-Regourd,* 2009). Currently this is only a hypothesis, and more observations are necessary to understand this effect; it is described here mainly to show the remote sensing power of the correlation between color and polarization.

Simultaneous observations of color and polarization and their combined analysis not only allow one to indicate and qualitatively explain changes in the particle properties, but also allow narrowing down the range of the particle parameters at computer modeling, making their results less ambiguous. An example of a comprehensive analysis that involves a simultaneous consideration of the color and polarization variations in the coma together with the changes in the brightness radial profile and Afρ in several filters is presented by *Ivanova et al.* (2019). The unusually deep negative polarization branch (down to –8%) for Comet C/2014 A4 (SONEAR) required a low-absorbing material; this, combined with the red color observed, brought tholin into consideration, which perfectly fit these requirements and was a realistic material for large heliocentric distances. The observed decrease in Afρ with the distance from the nucleus could be caused by decreasing either the albedo or size of particles, but simultaneous decrease in color and negative polarization ruled out particle fading and supported the idea of the dust fragmentation. Thus, the combination of color and polarization provided information about the composition of the dust in Comet C/2014 A4, and the combination of their variations with variations of other observables allowed revealing the evolution of the dust particles in the coma.

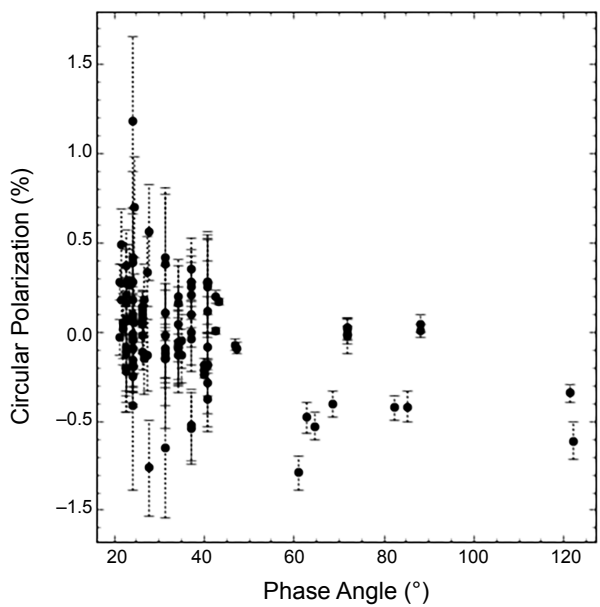

**Fig. 6.** Circular polarization data from *Kiselev et al.* (2017).

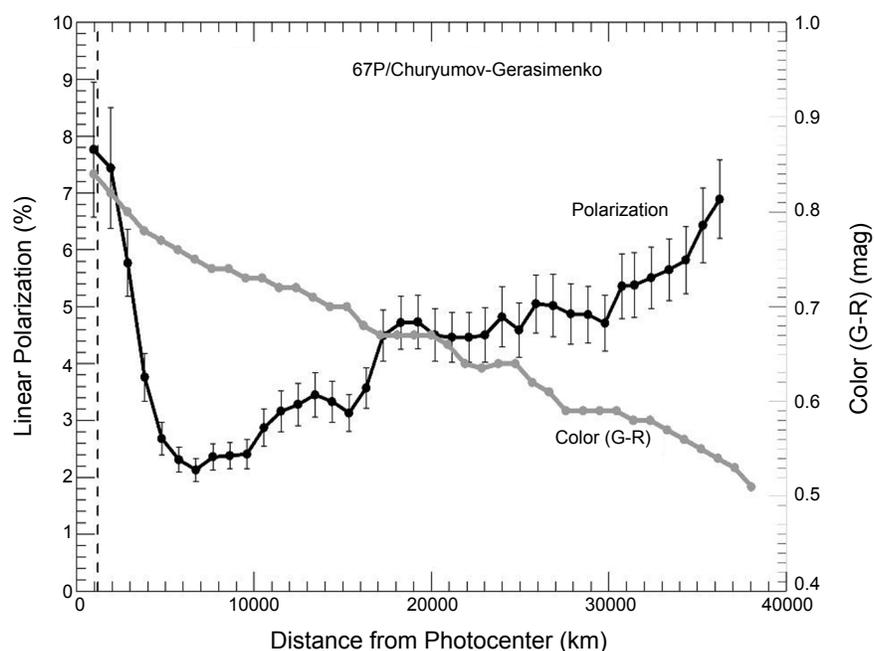

**Fig. 7.** Change of g-r$_{sdss}$ color and r$_{sdss}$ polarization of Comet 67P with the distance from the nucleus (*Rosenbush et al.,* 2017).

## 2.3. Near-Infrared Scattering and Mid-Infrared Emission

In contrast with optical brightness and color, NIR and MIR (~1–5 μm and ~5–40 μm respectively) wavelengths contain features that are diagnostic of ice and dust composition, although size and porosity also have their effects. Since the publication of *K_C2* and *Hanner and Bradley* (2004), NIR spectroscopy has become more common, and the Spitzer Space Telescope has afforded detailed MIR spectra of several comets.

*2.3.1. Near-infrared.* The NIR absorption bands due to water ice are located at 1.5, 2.2, and 3.1 μm. The 3-μm band is the strongest feature, but at wavelengths that are difficult to observe from groundbased observatories, due to strong telluric emission and absorption. After suggestive evidence of water ice in the 1980s, the first clear evidence (multiple features at high signal-to-noise ratios) was seen at Comet Hale-Bopp with spectra of the 1.5- and 2.0-μm absorption bands when the comet was at 7 au from the Sun (*Davies et al.,* 1997). Hale-Bopp's 3-μm band was subsequently observed at 2.9 au with the Infrared Space Observatory (ISO), followed by far-infrared emission features at 44 and 65 μm observed at 2.8 au from the Sun (*Lellouch*

*et al.,* 1998). Water ice detections in the NIR became more frequent in the new millennium, with several detections from the ground (*Kawakita et al.,* 2004; *Yang,* 2013; *Yang and Sarid,* 2010; *Yang et al.,* 2009, 2014; *Protopapa et al.,* 2018; *Kareta et al.,* 2021). Even though water appears to be the most abundant volatile in cometary nuclei, there are only a few direct detections of water ice in comae.

The relative shapes and strengths of the water ice absorption bands can be used to retrieve the properties of the ice grains, specifically, their size and chemical phase. Although the techniques to retrieve grain properties are not uniform, most investigations are consistent with micrometer-sized pure water ice grains (*Protopapa et al.,* 2018). The lack of a 1.5-μm feature in the ejecta of the mega-outburst of Comet 17P/Holmes and the subtle presence of the band at Comet C/2011 L4 (PanSTARRS) has been interpreted as indicative of submicrometer grains (*Yang et al.,* 2014). *Mastrapa et al.* (2008, 2009) showed that the NIR spectra have the potential to discriminate between ice in crystalline or amorphous forms. Particular emphasis has been given to a narrow feature at 1.65 μm that has only been observed following a large outburst of Comet P/2010 H2 (Vales) at 3.1 au from the Sun (*Yang and Sarid,* 2010). This feature is strongest in crystalline ice, but only for grain temperatures colder than ~200 K (*Grundy and Schmitt,* 1998). As a consequence of the similarity between warm crystalline ice and amorphous ice at 1.65 μm, but also due to the low crystallization temperature of amorphous ice (~160 K), amorphous water ice has not been definitively observed in a comet (*Kawakita et al.,* 2006, *Protopapa et al.,* 2014). For more on amorphous ice, see the chapter in this volume by Prialnik and Jewitt.

The variation of water ice features with time can also be used to infer coma ice grain properties. *Protopapa et al.* (2018) observed Comet C/2013 US$_{10}$ (Catalina) from 1.3 to 5.8 au. Spectra of the comet at rh ≥ 3.9 au displayed the 1.5- and 2.0-µm bands, but spectra at ≤2.3 au were featureless in this respect. The variation with heliocentric distance is consistent with the limited lifetimes of water ice in contact with low-albedo material. *Protopapa et al.* (2018) used effective medium approximation (see section 3.1) and an ice sublimation model to conclude that ice grains in the coma of C/2013 US$_{10}$ likely contain a small amount of dust, up to ~1% by volume.

**2.3.2. Mid-infrared (thermal) spectra.** At longer wavelengths, thermal emission dominates the spectra of comets. The thermal spectral energy distribution, defined by the temperature of the dust, was considered in detail in *K_C2*; here we focus on the studies where the main progress has been achieved: studies of the MIR spectral features.

The MIR silicate features arise from stretching and bending modes in Si–O bonds. The two main silicate types in comets are olivine and pyroxene and the spectral features in cometary comae are consistent with both classes in crystalline and amorphous phases. The "amorphous" materials may not be minerals in the strict sense, but appear to have stoichiometric compositions similar to olivine [Mg,Fe]$_2$SiO$_4$ and pyroxene [Mg,Fe]SiO$_3$ (see the chapter in this volume by Engrand et al.). The amorphous materials produce broad (≥1 µm) emission features at ~10 and ~20 µm. The shapes of the features vary with Fe-to-Mg ratio, and spectra of Comet Hale-Bopp were in best agreement with model spectra based on amorphous pyroxene and olivine with Fe/Mg = 1.0 (*Harker*, 1999). The crystalline minerals produce a variety of narrow (~0.1–0.5 µm) features throughout the MIR (*Dorschner et al.*, 1995; *Koike et al.*, 2000, 2003).

These features are most-consistent with Mg-rich crystalline species. A spectrum of Comet 17P/Holmes in Fig. 8 shows prominent peaks near 10.0, 11.2, 16.3, 19.3, 23.5, 28.0, and 33.5 µm arising from Mg-rich olivine. In contrast, crystalline pyroxene features tend to be weaker and not always well separated from Mg-rich olivine features, but evidence for Mg-rich crystalline pyroxene is found at 9.3 µm (*Wooden et al.*, 1997, 1999), ~14.4, ~15.4, and 29.4 µm.

Outside the silicate emission bands, there is a strong pseudo-continuum arising from carbon-bearing materials (see the chapter in this volume by Engrand et al.). The spectral temperature of this pseudo-continuum is up to ~30% warmer than a blackbody sphere at the same heliocentric distance (*Gehrz and Ney*, 1992; *Sitko et al.*, 2004). The continuum temperature is generally measured with a Planck function fit to data near 7.5 to 8.5 and 12 to 13 µm and the best-fit temperature normalized by the temperature of a large blackbody sphere (278 R$_h$$^{-1/2}$ K, for R$_h$ in astronomical units). The mean flux density of the spectrum in the middle of the silicate band (~10 µm) is normalized by the best-fit Planck function to derive the silicate band strength. *Gehrz and Ney* (1992) and *Sitko et al.* (2004) have shown that strong silicate features apparently correlate with continuum temperature. If the increase in silicate band strength was solely due to an increase in the relative abundance of silicate dust, then an anti-correlation between silicate feature strength and continuum temperature might be expected, as the long-wavelength continuum points (~12.5 µm) have silicate emission, whereas the short-wavelength points (~7.8 µm) have little to no emission, thus cooling the spectral temperature. Therefore, some other coma grain properties (e.g., size distribution or porosity) must be involved (see section 2.3.3).

At shorter wavelengths (3 to 5 µm), *Bockelée-Morvan*

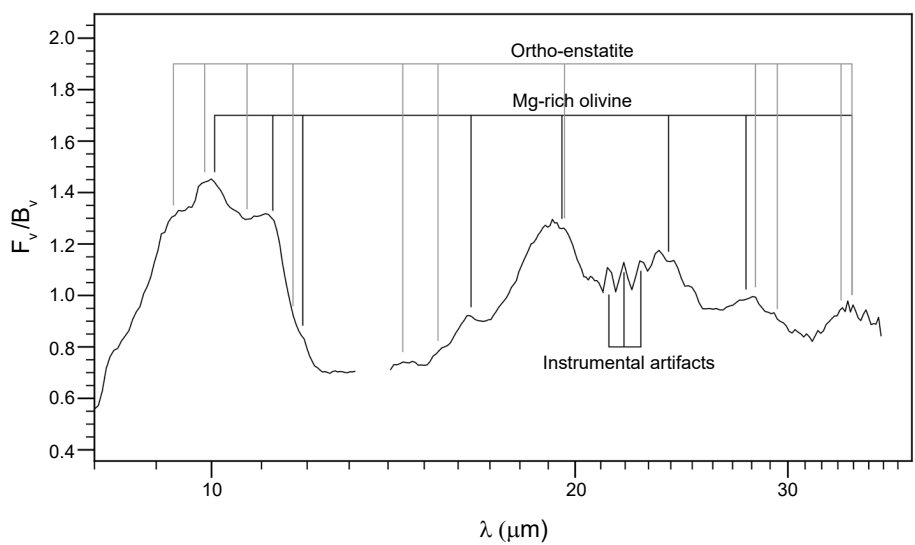

**Fig. 8.** Spectrum of Comet 17P/Holmes (*Reach et al.*, 2010; *Kelley et al.*, 2021) normalized by a Planck curve fit to the full wavelength range. The expected locations of strong features of Mg-rich olivine and orthoenstatite (Mg-rich pyroxene) from *Koike et al.* (2000, 2003) are marked. Also marked are selected weaker features of orthoenstatite at 14.4, 15.4, and 29.4 that may be present.



*et al.* (2019) found a correlation between continuum temperature (~0.3 K/° at 1.2 to 1.4 au) and phase angle in the inner coma of Comet 67P and identified thermal gradients supported by large grains (perhaps 100 μm or larger) as the likely cause. This effect has not been identified in groundbased observations but suggests that variations at the 10% level in coma spectral temperature may also be expected in the 10-μm region, depending on observing geometry and grain size distribution.

The presence of several other minerals has been considered based on cometary MIR spectra, and the broad spectral coverage afforded by spacebased or airborne observatories enabled these works. However, results claiming detections are lacking spectrally identifiable features. The challenge to confirming new minerals is identifying individual spectral features, rather than solely relying on the reduced χ-square fitting. The effects of grain shape are also important to consider, especially with anisotropic minerals like Mg-rich olivine (*Lindsay et al.,* 2013). *Lisse et al.* (2007) suggested evidence for carbonates in an ISO spectrum of Hale-Bopp. However, the feature at 7.0 μm is marginal (*Crovisier et al.,* 1997). Carbonates are not seen in other comets, and neither has the 12.7-μm feature been identified (*Woodward et al.,* 2007; *Bockelée-Morvan et al.,* 2009). To date, only the four silicate types (amorphous and crystalline pyroxene and olivine) remain as spectrally identifiable dust materials in MIR observations of comets.

### 2.3.3. Correlation between polarization and strength of the silicate feature in the thermal infrared.

The correlation between the values of maximum polarization and the strength of the 10-μm silicate feature was first considered by Hanner (2003). Lisse et al. (2002) and Sitko et al. (2004) attributed strong silicate features to small particles, whereas weak or absent silicate features were suggested as a signature of large particles. This sounds consistent with the polarimetric data, as small particles can be responsible for higher polarization due to higher contribution of the Rayleigh scattering, whereas the light scattered by large porous particles can experience a strong depolarization. However, this approach cannot explain the change in polarization with the distance from the nucleus (Fig. 5), as this would lead to an unrealistic conclusion that in the gassy comets, the dust is dominated by small particles near the nucleus and by large particles farther from the nucleus.

A more detailed analysis of this correlation was accomplished by *Kolokolova et al.* (2007). They showed that the correlation more likely results from different porosity of the dust particles. Their discrete dipole approximation (DDA) modeling (see section 3.1) of the silicate feature showed that for very fluffy particles, e.g., those presented by the ballistic-cluster-cluster aggregates (BCCA), the silicate feature remains strong no matter what size the particle (defined by the number of submicrometer monomers) is, whereas for less-porous ballistic-particle-cluster aggregates (BPCA), the silicate feature becomes weaker with increasing number of monomers (see more on BPCA and BCCA in section 3.2.1). Thus, one can expect that the dust in the comets with a

weak silicate feature is dominated by rather large compact particles, while in the comets with a strong silicate feature, the particles can be also large but more porous. At the same time, compact particles tend to be more gravitationally bounded to the nucleus and less accelerated by the gas flow than lighter porous particles of the same size that made them concentrated close to the nucleus. This leads to a decrease in the dust/gas ratio and increase in gas contamination at large apertures. As a result, comets whose dust is dominated by more compact particles (and show a weak silicate feature) should exhibit a decrease in polarization with an increase in aperture and overall lower polarization than comets with the dust dominated by more porous particles. Thus, the size of particles in all comets may be similar, dominated by large particles, but low-polarization and weak-silicate-feature comets are characterized by rather compact particles, whereas for the high-polarization and strong-silicate-feature comets, more porous particles are typical. The conclusion appeared to be consistent with the dynamical characteristics of the comets, specifically, the comets with supposedly more compact particles have smaller perihelia, and thus may be more affected by the solar radiation. The main conclusion of the paper by *Kolokolova et al.* (2007) was that two polarimetric classes of comets may result from a difference in the porosity of the dust particles.

These ideas have been supported by a more careful analysis by *Kwon et al.* (2021), who showed that the compositional differences in comets assumed based on the thermal infrared observations cannot explain the difference in cometary polarization, which is most likely associated with their different porosity.

## 3. INTERPRETATION TECHNIQUES

### 3.1. Most Popular Light-Scattering Techniques to Model Cometary Dust

For the cometocentric distances resolved by Earth-based observations, the number density of the cometary dust is low, which allows considering interaction of the dust particles with the solar radiation in the single scattering regime; i.e., one can ignore multiple scattering and other interactions between the particles and use computer models that consider light scattering by individual particles. Of course, one particle cannot represent cometary dust, which is characterized by particles with some size distribution of a variety of shapes and compositions. But combining simulations for different single particles, or even only for those which produce the dominating contribution to the scattered light, allows receiving a rather realistic model of the cometary dust.

Very active comets can have optically thick coma close to the nucleus (e.g., *Rosenbush et al.,* 1997). To simulate these observations, multiple scattering should be accounted for, and radiative transfer should be applied to the modeling of the internal coma of such comets. The ejecta produced by



Deep Impact were also optically thick, and radiative transfer was used to model the observed characteristics of the ejecta (*Nagdimunov et al.,* 2014).

The simplest particle used in cometary dust models is a homogeneous sphere. Its light-scattering characteristics (brightness, polarization, and their change with phase angle and wavelength) can be calculated using Mie theory (*Mie,* 1908), which represents an analytical solution of the Maxwell equations for the interaction of electromagnetic waves with a spherical particle. Numerous Mie codes are available online (see *https://scattport.org/index.php/light-scattering-software/mie-type-codes*) and are included in some computer libraries [e.g., interactive data language (IDL) routine Bohren and Huffman Mie (BHMIE) code]. Although it is known that cometary particles are not such spheres, this approach is still regularly used in cometary science, e.g., for interpretation of Rosetta data (e.g., *Fulle et al.,* 2010; *Fink and Rinaldi,* 2015; *Bockelée-Morvan et al.,* 2017), as it provides reasonable results when such properties as scattering and extinction cross-section are of interest (e.g., at computations of spectra) or when fundamental regularities in light scattering by particles are the subject of the study. Notorious resonance structures of Mie results (oscillations in the dependences of the light-scattering characteristics on phase angle or wavelength) can be smoothed out by using a broad size distribution of particles.

A more sophisticated but still a rather simple approach uses non-spherical regular particles, e.g., spheroids, cylinders, or other axisymmetric particles. For modeling such particles, there are direct theoretical solutions of Maxwell equations, reviewed in section 3.3.1 of *K_C2*, as well as solutions that use the T-matrix approach, described in *K_C2*, their section 3.3.2. Considering cometary dust as an ensemble of polydisperse and polyshaped (e.g., spheroids of different axes ratio) particles provides a rather realistic approach to the cometary dust (*Kolokolova et al.,* 2004b), especially if a model counts on the roughness of the particles [see the result for rough spheroids in *Kolokolova et al.* (2015)].

It is well known that cometary particles are not homogeneous, and a typical cometary particle represents an agglomerate of smaller grains (see the chapter in this volume by Engrand et al.). Although a realistic modeling of such particles requires the very sophisticated modeling tools reviewed in section 3.2, quite often such particles can be modeled using effective medium approximations (EMA), also called mixing rules. In EMA, an effective refractive index of a mixture of different materials (including voids) is considered based on the refractive index and volume fraction of each component in the mixture. The most popular mixing rules are those of Maxwell Garnett and Bruggeman [see a review of the various EMA in *Kolokolova and Gustafson* (2001); mixing rules were also considered in detail in section 3.2 of *K_C2*]. *Mishchenko et al.* (2014) showed that EMA can be derived directly from the Maxwell equations, thus representing a rigorous solution for interaction of electromagnetic wave with a media formed by a mixture of materials. *Mishchenko et al.*

(2016) also explored the limitations of the EMA, demonstrating that EMA works better for the smaller size parameter of the heterogeneities (inclusions) and for the number of inclusions per unit volume kept within some boundaries. The threshold value of the inclusion size parameter depends on the refractive-index contrast between the host and inclusion materials that often does not exceed several tenths, especially in calculations of the scattering matrix and the absorption cross section. Nevertheless, they showed that even for the materials with strongly contrasting refractive indexes such as hematite and air, there are quite realistic ranges of the sizes of the inclusions and host particle, and volume fraction of inclusions when EMA can be used safely. It is worth noting that one should be careful modeling layered particles using EMA; *Liu et al.* (2014) showed that EMA are reliable only for well-mixed cases down to an inhomogeneity scale, whereas the applicability of the EMA to cases of stratified or weak mixing materials is very limited. Thus, one should be careful using EMA to model porous particles in the visible; however, it is probably safe to use it for mixtures of silicates and organics (see section 3.2.2) or modeling cometary dust in thermal, or even NIR, wavelengths.

We would like to mention one more method that provides reasonable results, as it is an evident simplification of the realistic particles: distribution of hollow spheres (DHS) (*Min et al.,* 2005a). The DHS presents a rigorous solution of the Maxwell equations for coated spheres and considers the particles as a distribution of spherical shells characterized by the fraction, f, occupied by the central vacuum inclusion that changes from zero to some maximum value. Despite the departure of DHS from realistic dust particles, it models realistically looking spectra and even phase curves of polarization of laboratory samples (*Min et al.,* 2005a), allowing derivation of the particle size distribution. DHS was applied to comets only once (*Min et al.,* 2005b), and there are doubts that DHS can reproduce the properties of cometary agglomerates (*Levasseur-Regourd et al.,* 2020; *Tazaki and Tanaka,* 2018). However, this simple technique is extensively used to study other types of cosmic dust, specifically to model the 10-μm silicate feature in debris disks (e.g., *Arriaga et al.,* 2020).

A more comprehensive study of the light scattering by realistic cometary particles can be achieved by solving the Maxwell equations after some adjustments that allow accounting for the shape and structure of complex scatterers. Numerous techniques that make the Maxwell equations solvable for complex particles were reviewed in *K_C2*. Among them, there are two techniques most often used for modeling cometary dust: the T-matrix approach and coupled dipole approximation (more often called discrete dipole approximation, or DDA). The popularity of those techniques is partly related to the fact that T-matrix and DDA codes have been available online for many years; they are regularly updated and are accompanied by a detailed documentation. The main ideas behind those techniques were considered in *K_C2*, so here we focus only on the new developments of these techniques and their application to cometary dust.



For a more detailed review of those techniques, as well as other light-scattering approaches to the light scattering by particles, see *Kimura et al.* (2020).

The main idea of the DDA is that a particle is divided into small cells, each of which is considered as a dipole. The DDA technique yields a system of linear equations that describes the fields that excite each dipole: the external field and the fields scattered by all other dipoles. The numerical solution of these equations allows accounting for the contribution of all dipoles in the total scattered field produced by the particle. The main advantage of the DDA approach is its flexibility; it can be applied to particles of arbitrary shape, structure, and composition. See the most popular discrete dipole scattering (DDSCAT) code and its detailed description at *http://ddscat.wikidot.com* and the latest DDSCAT user guide at *https://arxiv.org/abs/1305.6497*. There are two versions of the DDA codes in the DDSCAT package. The newer one utilizes fast Fourier transformation (FFT), which accelerates the computations but significantly increases the required RAM; the code without FFT is slower but requires less computing resources. We also recommend a powerful DDA package, called ADDA (*Yurkin and Hoekstra,* 2011), located at *https://github.com/adda-team/adda*.

To achieve an accurate solution, the number of dipoles should exceed $60|m-1|^3(\Delta/0.1)^{-3}$ (*Draine,* 1988), where m is the particle material refractive index and $\Delta$ is the fractional error (the ratio of the error to the quantity being computed). This puts significant requirements on computer resources. Besides, each DDA solution considers only one orientation of the particle, and to calculate a realistic randomly-oriented particle, the computations need to be done for numerous orientations whose number increases with increasing the size of particle and complexity of its structure. Note that some orientationally averaged DDA solutions can be obtained analytically — this and other ways to increase the DDA efficiency are considered in *Kimura et al.* (2016). The DDA technique has been especially successful in modeling thermal infrared spectra; see a review of those modeling efforts in the chapter in this volume by Engrand et al.

Unlike DDA, which provides the solution considering the internal fields of the scatterer, the T-matrix method is based on a solution to the boundary conditions on the particle surfaces and was originally called "extended boundary condition method" (*Waterman,* 1965). The method expands the incident and scattered fields into vector spherical functions with the scattered field expanded outside a sphere circumscribing a non-spherical particle. A significant development of the method was an analytical orientation-averaging procedure (*Mishchenko,* 1991) that made calculations for randomly oriented particles as efficient as for a fixed orientation of the same particle. It has been applied to light scattering by spheroids, cylinders, and Chebyshev particles (*Mishchenko et al.,* 2002); the codes can be downloaded from *https://www.giss.nasa.gov/staff/mmishchenko/tmatrix/*.

In the case of modeling cometary dust, the most popular development of the T-matrix technique is the superposition T-matrix method, which allows computing the light-scattering characteristics of clusters of spheres (i.e., agglomerated particles), calculating the T-matrix for each monomer in the cluster and then calculating the T-matrix for the whole cluster summarizing the scattered external field from other spheres in the cluster. To some extent, it uses the idea of the DDA; however, instead of dipoles, it considers the T-matrixes of individual monomers. The superposition approach allows applying the main T-matrix advantage, namely, analytical averaging over orientations. However, it has been recently showed that in some cases it is more efficient to average over orientations numerically after computations over a set of orientations is done. The main advantage of this type of computation is that it requires less RAM and thus allows considering larger clusters. The most recent version of the superposition T-matrix code, called the multiple sphere T-matrix (MSTM), is available at *https://github.com/dmckwski/MSTM*. It allows considering the clusters formed by spheres of different sizes and compositions, which can be arranged inside or outside of other spheres, thus modeling layered spheres or spheres with spherical inclusions, and/or adjacent to multiple plane boundaries, and/or in two-dimensional periodic lattices. It also allows modeling clusters made of linearly and circularly birefringent and dichroic materials and thus can be applied to crystals and homochiral (e.g., biological) particles (see *Nagdimunov et al.,* 2013).

A significant progress in modeling large particles has been recently achieved with a new development in the T-matrix approach called the fast superposition T-matrix method (FaSTMM), introduced by *Markkanen and Yuffa* (2017). The FaSTMM uses the fast multipole method (FMM) to speed up the STMM solution (which is similar to the MSTM considered above). The FMM forms monomer groups hierarchically and computes electromagnetic interactions between the separate groups in each level of hierarchy. This decreases the costs of computing all the pairwise monomer interactions in a system of N monomers from $O(N^2)$ to $O(N *\log N)$. Note that like the DDA, specifically the DDSCAT code, the FaSTMM does not allow analytical orientational averaging. The FaSTMM code is available at *https://wiki.helsinki.fi/xwiki/bin/view/PSR/Planetary%20System%20Research%20group/*. FaSTMM was successfully used to model Rosetta dust particles (*Markkanen and Agarwal,* 2019); for more information, see section 3.2.1.

One more important development in the light-scattering simulations is the approach to model light scattering by particulate surfaces using the so-called radiative transfer-coherent backscattering (RT-CB) code. Although the approach contains some approximations, it provides quite realistic results on light scattering by particulate surfaces, including such coherent backscattering effects as the opposition spike and negative polarization spike observed at very small phase angles (e.g., *Mishchenko et al.,* 2009). This technique considers a particulate surface as a sparse layer of spheres and is based on the Monte Carlo vector radiative transfer code wherein coherent backscattering is computed by incorporating the reciprocity relation in electromagnetic



scattering and keeping the relative phase information of the wave components (*Muinonen et al.*, 2012, 2015). Strong coherent backscattering effects were not observed for cometary surfaces (*Masoumzadeh et al.*, 2019), so the RT-CB code was not used in modeling cometary nuclei. However, a version called radiative transfer with reciprocal transactions ($R^2T^2$) was successfully combined with FaSTMM (*Muinonen et al.*, 2018; *Markkanen et al.*, 2018a). Unlike RT-CB, $R^2T^2$ is used for densely packed layers and incorporates the full incoherent extinction, scattering, and absorption properties of a volume element larger than the wavelength. The RT-CB code requires particles to be spherical because it uses the Mueller matrix to compute the scattering interactions. The Mueller matrix does not contain electromagnetic phase information (required for the coherent backscattering CB part) except for spherical particles. $R^2T^2$ code is basically RT-CB code but the scattering interactions are computed using the T-matrix instead of the Mueller matrix. The FaSTMM solution combined with the $R^2T^2$ approach was successfully applied to model Rosetta dust particles by *Markkanen et al.* (2018b). For more details and comparison of the different techniques described above, see *Penttilä et al.* (2021).

The other important source of information about light-scattering characteristics of cometary particles is laboratory measurements of cometary dust analogs. Systematic laboratory simulations of light-scattering characteristics of dust particles are performed by four main groups: the Granada-Amsterdam group, specialized in measuring light-scattering characteristics of cosmic analogs; the Paris group, especially known by their low-gravity experiments; and the Bern and Grenoble groups, which perform light-scattering experiments with ice and its mixtures with other materials. A review of the laboratory measurements performed by the Bern and Grenoble groups is presented in the chapter in this volume by Poch et al. The Granada-Amsterdam group has produced a popular database of light-scattering properties of mineral and meteoritic particles (see *https://old-scattering.iaa.csic.es/*). This database provides a lot of photometric and polarimetric phase curves to compare with the astronomical observations described in section 1; they also measured a complete Mueller matrix of the samples that is intensively used to test theoretical tools (see review in *Muñoz and Hovenier*, 2015). The Paris group has collected a large number of the phase curves and wavelength dependences of polarization for particles analogous to the ones expected in comets and other cosmic bodies. This allowed them to find out the main regularities between the polarimetric characteristics of particles and their size and structure (e.g., *Levasseur-Regourd et al.*, 2015), especially in the case of complex particles, which is hard to model theoretically. It is also worth mentioning studies by groups that provide optical constants (complex refractive index) of the materials of astronomical interest. The data for various silicates, oxides, sulfides, and carbonaceous materials produced by the Jena group are presented in *https://www.astro.uni-jena.de/Laboratory/Database/jpdoc/f-dbase.html*. A compilation of several databases of optical constants can be found at the Jena-St. Petersburg Database of Optical Constants (HJPDOC) (*https://www2.mpia-hd.mpg.de/HJPDOC/database.php*), which still contains the largest collection of optical constants of materials of astronomical interest even though it has not been updated since 2008. The MIR refractive indexes for astronomical ices are produced by the Universidade do Vale do Paraíba (UNIVAP), Laboratório de Astroquímica e Astrobiologia (LASA), and can be found at

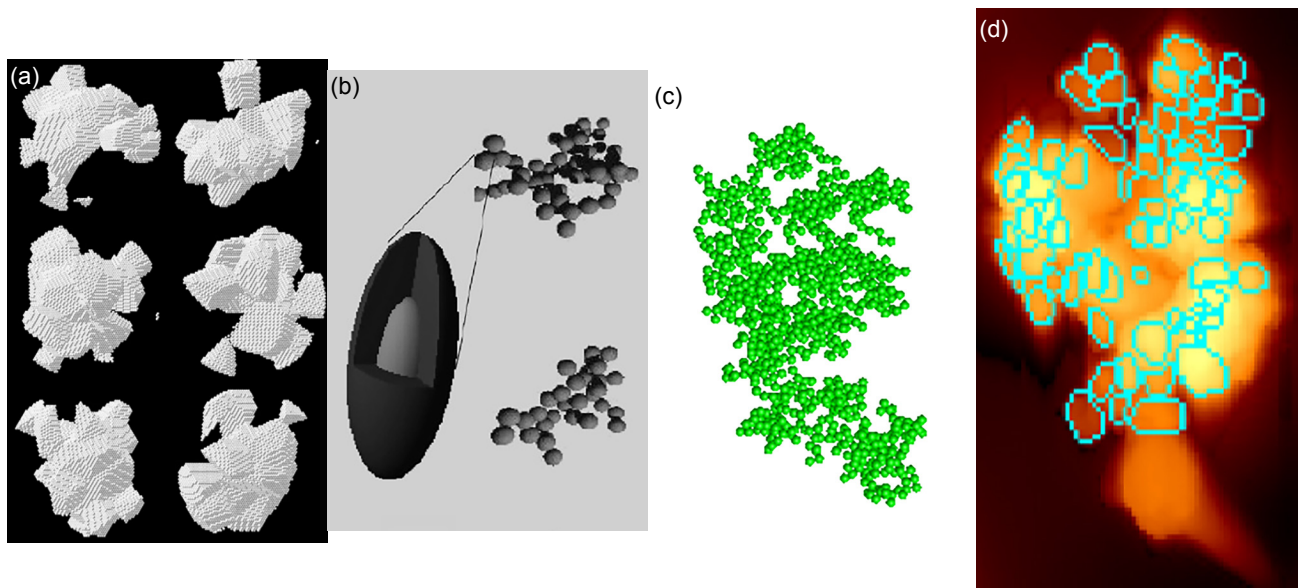

**Fig. 9.** Comparison between cometary particles used in numerical simulations and a real one. **(a)** "Agglomerated debris" modeled with DDA (*Zubko*, 2013). **(b)** Agglomerates of layered spheroidal monomers modeled with DDA (*Lasue et al.*, 2009). **(c)** Hierarchically structured cluster of 64 × 16 spheres modeled with MSTM (*Kolokolova et al.*, 2018). **(d)** Real cometary dust particle, imaged by Rosetta atomic force microscope MIDAS (*Mannel et al.*, 2019).



https://sites.google.com/view/astrow-en/research/databases. Some collections of optical constants can be found in the NASA PDS, e.g., for water ice by *Mastrapa* (2020).

## 3.2. Applications of Light-Scattering Techniques to Cometary Dust

*3.2.1. Size and structure of particles.* As was mentioned in the previous section, for interpretation of the remote sensing data on cometary dust, two main techniques, DDA and T-matrix, were used. After a success of modeling the cometary dust as porous aggregates to explain cometary photometric and polarimetric phase curves in Kimura et al. (2003a, 2006), the majority of the interpretation attempts focused on porous and aggregated particles. Figure 9 presents examples of the particles used to model cometary dust with DDA (Figs. 9a,b) and T-matrix (Fig. 9c); Fig. 9d shows an image of a real cometary particle.

The "agglomerated debris" shown in Fig. 9a have been extensively used in the papers by Zubko et al. since the publication of *Zubko et al.* (2005). They are produced by removing cells from a spherical particle, thus creating particles and monomers of irregular shapes, which is the main advantage of the model. However, to achieve a larger size of particles, the model increases the size of the monomers, keeping the larger particles structurally identical to smaller particles; i.e., they have the same number, shape, and position of the monomers, which are scaled to the new size. In reality, larger cometary agglomerates are characterized by a larger number of the monomers of the same size. Thus, Zubko's model does not provide information regarding the size of the monomers and cannot correctly reproduce the electromagnetic interaction between the monomers in realistic particles, which contain hundreds of monomers. As a result, the model attributes the observed differences in light-scattering characteristics for different comets primarily to a difference in the composition of their dust. This makes the model not only too rigid, but also leads to doubtful results.

A more realistic structure for cometary particles is offered by ballistic aggregates, and the following ballistic aggregates are most often used:

- BPCA that have a porosity of about 85% and in the case of large particles approach a fractal dimension of 3;
- BCCA that have porosity larger than 95% and in the case of large particles approach a fractal dimension of 2;
- Ballistic Agglomeration with One Migration (BAM1) and Ballistic Agglomeration with Two Migrations (BAM2) particles are characterized by porosities about 75% and 65% respectively; see *Shen et al.* (2008, 2009) for details.

A detailed description of the particle models and the techniques used for their generation are discussed in *Kimura et al.* (2020).

Closer to the realistic cometary particles is the particle in Fig. 9b where the spheroidal core-mantle monomers are organized in a ballistic aggregate (*Lasue et al.,* 2009). This model considers a size distribution of monomers as well as a size distribution of particles. However, due to a limitation

of computer resources, the model was limited to aggregates of 256 monomers, and modeling larger particles required a smaller number of larger monomers.

The particle in Fig. 9c is built of spherical monomers, which is the only opportunity in MSTM modeling. Although these particles are formed by monodisperse monomers, they may better represent a structure of the cometary agglomerates as they are hierarchical BPCAs. To create a hierarchical BPCA, regular BPCA are built and then these clusters are ballistically organized in the clusters of the second order, and then the clusters of each next order are built ballistically from the clusters of the previous order (*Kolokolova et al.,* 2018). Hierarchical structure is confirmed for the dust agglomerates in Comet 67P (*Mannel et al.,* 2019). This model is also consistent with the dust formation modeling in protoplanetary nebulae (*Dominik,* 2009).

*Kolokolova and Kimura* (2010b) presented cometary dust as a mixture of aggregated and solid non-spherical particles. *Lasue et al.* (2007) used a similar model for modeling interplanetary dust. Lasue et al. used DDA in their modeling, and Kolokolova and Kimura used the MSTM code for aggregates and the T-matrix code for spheroids for compact particles. In both studies the mixture of agglomerated and compact particles allowed the reproduction of brightness and polarization phase curves. The model by *Kolokolova and Kimura* (2010b) also reproduced red color and polarimetric color of the dust and its low albedo. The ratio of compact to fluffy particles appeared to be close to the one found *in situ* for Comet Halley, and the mass ratio of silicate to carbonaceous materials equal to unity in accordance with the elemental abundances found by the Giotto mission in Comet Halley (e.g., *Jessberger et al.,* 1988).

Probably the most comprehensive model of the cometary dust has been presented by *Halder and Ganesh* (2021), in which a mixture of porous, fluffy, and solid (according to the definitions from *Güttler et al.,* 2019) particles was considered. They created high-porosity hierarchical (HA) particles and moderate porosity structures with solids in the core, surrounded by fluffy aggregates called fluffy solids (FS). They also added solids as low-porosity (<10%) particles in the form of agglomerated debris. They studied the mixing combinations: HA and Ssolids; HA and FS; and HA, FS, and solids. Complexity of the model and the efforts of the authors to bring the model close to the properties of cometary dust, as they are known from the Rosetta studies, are very impressive. However, this work also shows a limitation of such complex multi-parameter models: The fit to polarization phase curve for a specific comet (the paper considered Halley, 67P, Hyakutake, Hale-Bopp) could be achieved with different compositions of the particles, depending on their size distribution and the ratio of fluffy to solid particles in the mixture. The authors tried to narrow down the range of the best-fit particle characteristics, adding to the analysis the spectral polarimetric data. They presented a set of the unique particle characteristics that provided the best fit to both angular and spectral polarimetric data. However, a reliability of these unique characteristics is not evident. For example,



neither the composition of the Comet 67P dust (50% silicates and 50% carbon) nor the ratio of porous to solid particles (75% solids) are consistent with the Rosetta data (*Bardyn et al.,* 2017; *Güttler et al.,* 2019), and the size distribution (power –2.4) is not consistent either with the Grain Impact Analyser and Dust Accumulator (GIADA) data (*Fulle et al.,* 2016) or with Visible and Infrared Thermal Imaging Spectrometer (VIRTIS) data (*Rinaldi et al.,* 2017), as they derived the power >3 in absolute value for the dates close to the observations of 67P by *Rosenbush et al.* (2017) used for fitting by Halder and Ganesh. One more reason why the obtained best-fit parameters may not reproduce the realistic properties of the dust particles is that the considered particle was <15 μm in size, whereas an abundance of much larger particles was detected by the Rosetta mission and in the tail of Comet 67P (*Fulle et al.,* 2016; *Moreno et al.,* 2017).

The main problem with modeling realistic cometary particles is that their size may reach hundreds of micrometers and even millimeters (*Güttler et al.,* 2019). Unfortunately, the current computer resources limit capabilities of DDA or T-matrix computations, which require enormous RAM and many hours of computation time even for the case of parallelized version of the codes. So far, the largest particles computed with DDA were ~10 μm in size (e.g., *Zubko et al.,* 2020); note that they were agglomerated debris formed just by a dozen of micrometer-sized monomers. The largest particles modeled with the MSTM code are also about 10 μm, consisting of ~2000 submicrometer monomers (*Kolokolova and Mackowski,* 2012). Recent versions of the MSTM code whose efficiency is dramatically increased with the FFT procedure (*Mackowski and Kolokolova,* 2022) allowed computing particles up to 16,000 submicrometer monomers, i.e., reaching particle sizes about 20 μm. However, this is still smaller than the particles studied by the COmetary Secondary Ion Mass Analyzer (COSIMA) and GIADA instruments on the Rosetta spacecraft.

The largest particles, up to 100 μm, were modeled by *Markkanen et al.* (2018b). Their model considered a volume shaped as a Gaussian (see *Muinonen et al.,* 1996) particle, filled with a mixture of submicrometer organic and submicrometer-sized silicate spherical monomers. The monomers were polydisperse, covering the range of radii from 50 to 1000 nm. *Markkanen et al.* (2018b) considered a power-law size distribution of those particles with the power –3 and varied the number of organic and silicate monomers in the mixture as well as the volume fraction of the voids in the particle. The light scattering by such a particle was modeled using the FaSTMM and $R^2T^2$ techniques described in section 3.1. The best fit to the Rosetta photometric phase curve was found for the dominant particle size between 5 and 100 μm. The total volume of the organic grains was estimated as 0.3 and that of silicate grains estimated as 0.0375, which, being converted to the mass fraction, is reasonably consistent with COSIMA (*Bardyn et al.,* 2017) estimations. As the authors noted, the presence of silicate monomers with organic mantles can easily increase the abundance of the silicates in the dust.

*Markkanen et al.* (2018b) also modeled the nucleus phase function, presenting the nucleus surface as an ensemble of 1-km-sized Gaussian particles filled with smaller particles. This allowed them to introduce a surface roughness that followed Gaussian statistics, resulting in the corresponding geometric shadowing effect. The results showed that the nucleus could be simulated with the same particles as in the coma but packed densely on a rough surface. Although the main goal of this paper was to model Rosetta's Optical, Spectroscopic, and Infrared Remote Imaging System (OSIRIS) phase curve of the coma and nucleus, photometric, and polarimetric phase curves observed from the Earth are also among the results and are presented by the particles of size <20 μm. This provides strong hope that fitting groundbased observations does not require particles larger than some tens of micrometers. Probably the only questionable feature of this model is that the particle is presented as a volume filled with constituents that are not in direct contact with each other. As was shown in the laboratory measurements with icy particles (see the chapter in this volume by Poch et al.) and in *Kolokolova and Mackowski* (2012), the light-scattering characteristics of the particles assembled in some structures is different from the freely distributed, not connected, particles, probably due to stronger near-field effects.

Thus, all currently used models have some significant limitations that may affect the trustworthiness of the applications to specific comets. However, the developed models can be successfully used to understand the physics of the interaction of the dust particles with radiation and to uncover how the physical properties of the particles affect the observed characteristics. Specifically, one can study dependences of brightness and polarization on the size and composition of monomers and their arrangement in a particle using highly controlled models such as ballistic aggregates. Examples of this type of study can be found in *Kimura et al.* (2006), *Shen et al.* (2009), *Kolokolova and Mackowski* (2012), and *Kolokolova et al.* (2018). Recently *Mackowski and Kolokolova* (2022) have extended such modeling to large BPCAs reaching more than 16,000 monomers. The computations were made using the MSTM code upgraded with FFT capabilities and performed at the NASA High-End Computing Capability (HECC) facility. The results are shown in Fig. 10.

The plots in Fig. 10 allow us to draw some important conclusions. The monomer size (size parameter) strongly affects the polarization and brightness phase curves even if a very narrow range of values is considered. By increasing the number of monomers in the agglomerate, the polarization tends to reach some limit and does not change after the limit is reached. The number of monomers that defines the limit is larger for transparent materials. Thus, it may be plausible to avoid modeling agglomerates of more than 1000 monomers in the case of absorbing materials.

The most interesting aspect is the dependence on the composition. One can see that the polarization maximum is very high for ice, then decreases for silicates, then increases



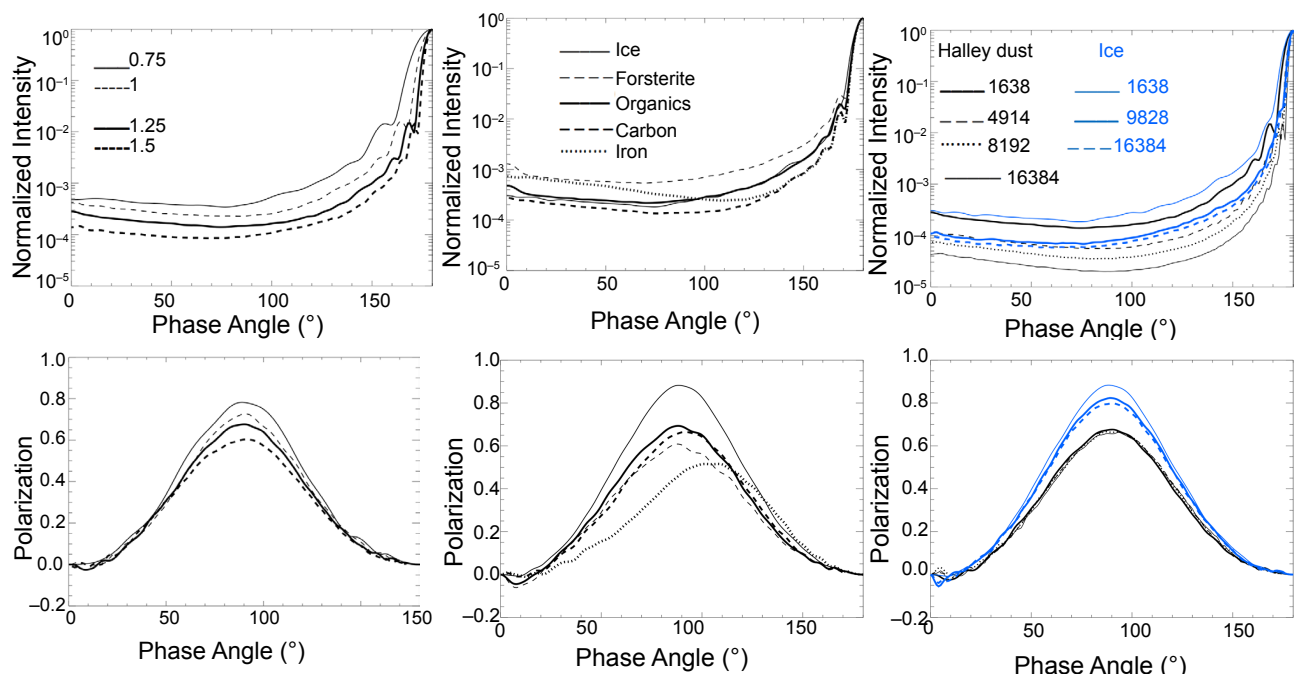

**Fig. 10.** Modeled phase curves of normalized brightness (top panel) and polarization (bottom panel) for large clusters of particles from *Mackowski and Kolokolova* (2022). *Left column:* Effects of the monomer size parameter for Halley dust (see section 3.2.2). *Middle column:* Results for different materials. *Right column:* Effects of the number of monomers for Halley dust and ice. Thicker font in the legend indicates thicker line. If not specified, the monomers of radius 0.1 μm and agglomerates of 1638 monomers were used. In all cases the porosity of the agglomerates was ~85% and the wavelength was 0.65 μm.

for more absorptive materials like Halley dust (see section 3.2.2) and carbon, but then strongly decreases for even more absorptive materials such as Fe. Thus, the dependence on the absorption (imaginary part of the refractive index) is not straightforward and can result in an increase or decrease of polarization.

The observed regularities can be explained from the point of view of the polarizability of the particle units. The basic polarizability, α, is polarizability of the molecules in the particle material related to the material refractive index through the Lorentz-Lorenz equation $\alpha \sim (m^2 - 1)/(m^2 + 2)$. Thus, in the case of a low imaginary part of the refractive index, polarizability is larger for a larger real part. However, the situation is more complex for agglomerates as stronger polarizability results in stronger interactions between the monomers, causing stronger depolarization of the light. As a result, icy clusters with a low refractive index have weakly interacting monomers, which explains why positive polarization produced by an icy agglomerate is larger than that for silicates. For a larger imaginary part of the refractive index, the polarizability is defined by a complex interplay between the real and imaginary parts of the refractive index and polarization may increase or may decrease (see Fig. 10). More examples can be found in *Kimura et al.* (2006). There, in Fig. 1, for the same imaginary part of the refractive index, polarization of aggregates decreases with an increase in the real part. In Fig. 2 of the same paper, for n = 1.4–1.6, the maximum polarization decreases with an increase in the real part, but the trend is opposite for n = 1.7–2.0.

We also see in the top panel of Fig. 10 a strong dependence on the monomer size. The reason for this can again be in the polarizability, but now in the polarizability of a sphere in the cluster, which is defined by the total number of dipoles (molecules) in the sphere and thus increases for larger monomers consisting of more dipoles. Monomers with larger polarizability interact more strongly, and the effects of electromagnetic interaction produced by them more greatly depolarize the light.

### 3.2.2. Composition of the cometary dust.

Besides the size and structure of the cometary dust particles, an important parameter of the modeling is the particle composition. The chapter in this volume by Engrand et al. considers the composition of cometary dust in detail; this chapter focuses on how the composition of the dust is reflected in the refractive index that is used in light-scattering modeling.

It is now accepted that the main components of cometary dust are carbonaceous materials and silicates. For a long time, it was considered that the most abundant silicates in cometary dust are amorphous silicates, called glass with embedded metal and sulfides (GEMS), and to a lesser extent, anhydrous Mg-rich minerals like forsterite and enstatite (*Hanner and Bradley,* 2004), although *Stenzel et al.* (2017) showed that in the case of Comet 67P the Mg to Fe ratio was close to that of meteorites. The composition of the organics is more obscure. Previously, it was considered that the optical properties of cometary organics can be presented



as a mixture of amorphous carbon and so-called cosmic organics (*Li and Greenberg,* 1997). Rosetta findings indicate a material more similar to insoluble organic matter (IOM) although less processed than the IOM found in carbonaceous meteorites (*Paquette et al.,* 2021).

It is still a question how these materials are distributed in the particle. Based on Giotto's dust impact mass spectrometer, PUMA, the IDPs, and Rosetta COSIMA studies, it is very unlikely that cometary particles are pure silicate and pure carbonaceous particles forming two separate populations of the dust particles. Most likely, the silicate grains are embedded in the organic matrix or represent small silicate particles with organic mantles glued together. This allows considering mixtures of small silicate and organic monomers in a single particle (*Markkanen et al.,* 2018b) or a particle composed of an intimate mixture of silicate and metals encased in carbonaceous materials (*Mann et al.,* 2004) as more realistic models. The latter allows applying the EMA (see section 3.1) to calculate the refractive index of the particle.

*Mann et al.* (2004) suggested a so called "Halley-like composition" (aka Halley dust) that is an intimate mixture of the materials consistent with the elemental composition of comet dust measured by the mass spectrometer on the Giotto mission (*Jessberger et al.,* 1988). The refractive index based on this composition was successfully used to model cometary light-scattering characteristics in numerous papers. However, it may need to be updated using the *in situ* data for Comet 67P and the elemental composition of the dust consistent with the reported in *Stenzel et al.* (2017), *Bardyn et al.* (2017), and *Paquette et al.* (2021).

To derive the refractive index for the cometary dust material, in both cases of the Halley-like and 67P-like compositions, the silicates are presented by so-called astronomical silicate, an analog material of interstellar silicate with optical properties consistent with those for MgFeSiO$_4$ (e.g., *Laor and Draine,* 1993), which was successfully used to model light scattering by different types of cosmic dust. The organic material was selected based on the results of *Kimura et al.* (2020), who showed that cometary organic matter should be carbonized after the formation of comets, thus acquiring the elastic properties similar to those of hydrogenated amorphous carbon; then a collisional velocity of a few meters per second could result in the easy fragmentation of the dust aggregates, seen in COSIMA's optical microscope (COSISCOPE) images. A lack of the optical constants for carbonized organics forced us to assume that the refractive indexes of the hydrogenated amorphous carbon likely mimic those of carbonized organic matter, because carbonization is characterized by the loss of H, N, and O (*Jenniskens et al.,* 1993). Table 1 presents the optical constants of the materials used to derive the refractive indexes for cometary dust. The ratio of different components in the mixture was chosen to be consistent with the elemental abundances discussed above.

One can see that despite some differences in the composition of those comets, the refractive index and its spectral change are similar, and thus the light-scattering modeling obtained with Halley-like composition is relevant to the dust in 67P and can be recommended for use in light-scattering models.

## 4. PHYSICAL PROCESSES CAUSED BY INTERACTION OF THE DUST WITH ELECTROMAGNETIC RADIATION

### 4.1. Radiation Pressure: Physics and Dependence on the Properties of Particles

The interaction of electromagnetic waves with dust particles exerts a force on the particles approximately in the direction of wave propagation, known as radiation pressure (*Burns et al.,* 1979). When dust particles are exposed to solar radiation, the solar radiation pressure pushes the particles outward in the radial direction, i.e., the anti-direction to solar gravity. Because both the solar radiation pressure and the solar gravity are proportional to the inverse square of the distance from the Sun, the ratio β of solar radiation pressure to solar gravity is a non-dimensional quantity useful for studying the relative importance of radiation force to the dynamics of dust particles.

Dust particles in the vicinity of their parent body are also driven away from the surface of the body by radiation pressure due to the reflection of solar radiation and the thermal emission from the body (*Burns et al.,* 1979; *de Moraes,* 1994; *Bach and Ishiguro,* 2021). For the particles moving with respect to the Sun, a relativistic drag force, known as the Poynting-Robertson (P-R) effect, appears in the equation of motion in the direction against particle motion to terms on the order of v/c where v and c denote the velocity of the particle and the speed of light, respectively (*Robertson,* 1937). Owing to the proportionality of the P-R drag to the β ratio, the larger the β ratio of a particle is, the shorter the P-R lifetime of the particle is. As a result, only large particles with small β values can remain near the orbits of their parent bodies, which have been observed as dust trails and meteor showers.

The solar radiation pressure acting on a moving particle differs from that on a particle at rest in the reference frame of the Sun due to the Doppler effect on the radiation pressure cross section, although the effect is typically negligible (*Kimura et al.,* 2002a).

The direction of solar radiation pressure is not exactly radial for non-spherical particles, because scattering and absorption of solar radiation is in general asymmetric around the radial vector (*van de Hulst,* 1957). The radial and non-radial components of radiation pressure on non-spherical particles have been computed by the method of separation of variables for spheroids (*Il'in and Voshchinnikov,* 1998) and the a1-term method for fluffy aggregates (*Kimura et al.,* 2002a). The computations have shown that the non-radial radiation pressure on non-spherical particles is on average a tenth or hundredth of the radial component, although there is a specific orientation of the particles where the non-radial components are comparable to the radial component.



TABLE 1.  Refractive indexes of the materials and the effective refractive index for the dust in Comets Halley and 67P.

| Material name | Wavelength | | Volume fraction | | Reference |
|---|---|---|---|---|---|
| | 450 nm | 650 nm | 67P | Halley | |
| Amorphous carbon* | 1.95 + 0.786i | 2.14 + 0.805i | 0.5042 | 0.4379 | *Rouleau and Martin* (1991) |
| Organic refractory** | 1.69 + 0.150i | 1.71 + 0.149i | 0.2521 | 0.2189 | *Li and Greenberg* (1997) |
| Astronomical silicate | 1.69 + 0.0299i | 1.68 + 0.0302i | 0.2208 | 0.3176 | *Laor and Draine* (1993) |
| Iron | 2.59 + 2.77i | 2.90 + 3.02i | 0.0228 | 0.0256 | *Johnson and Christy* (1974) |
| Comet Halley | 1.88 + 0.47i | 1.98 + 0.48i | | | *Kimura et al.* (2003a) |
| Comet 67P | 1.901 + 0.526i | 2.015 + 0.532i | | | |

\* Represents carbonized organics.
\*\* Represents primordial organics.

The radial component of solar radiation pressure on fluffy aggregate particles has been computed using the various EMAs, the DDA, the generalized multisphere Mie solution (GMM), and the MSTM [see section 3.1 and review in *Kimura et al.* (2016)]. Figure 11 depicts the β ratios for spheres, BPCAs, and BCCAs consisting of 0.1-μm-radius monomers of amorphous carbon AC1, astronomical silicate, Mg-rich olivine, Mg-rich pyroxene, Mg-rich silicate with Fe inclusions, and silicate-core + organic-mantle, which are a compilation of *Kimura et al.* (2002b, 2003b), *Köhler et al.* (2007), and *Kimura* (2017). The β ratio has a maximum in the submicrometer size range, while the maximum ratio is smaller for fluffy agglomerates than for compact particles. In contrast, values of β for large fluffy agglomerates in the geometrical optics regime are higher than those for compact particles of equal mass, because the geometrical cross section of particles increases with fluffiness. While the maximum β for fluffy agglomerates depends on the size of monomers, the ratios for any pure silicate particles do not exceed unity, irrespective of the particle morphology (*Mukai et al.,* 1992; *Kimura et al.,* 2002b). However, if metallic Fe inclusions are embedded in a Mg-rich silicate matrix, then the maximum values of β could exceed unity (*Altobelli et al.,* 2016; *Kimura,* 2017). The values of β for Mg-rich silicate particles with metallic Fe inclusions are barely distinguishable from those of silicate-core + organic-mantle particles. While solar radiation pressure on particles composed of Fe-rich, Mg-poor silicates is higher than that on particles of Fe-poor, Mg-rich silicates owing to the effect of Fe atoms on the absorption efficiency in the visible wavelength range, solar gravity on the former is also higher than the latter due to the effect of Fe atoms on the density. Accordingly, one should not expect that the ratios of solar radiation pressure to solar gravity for Fe-rich, Mg-poor silicates are significantly different from those for Fe-poor, Mg-rich silicates.

As the particles move away from the sunlit surface, they are gradually decelerated by the solar radiation pressure. Based on Fig. 11, it is reasonable to assume the β ratios of organic-rich and silicate-poor (CHON) particles are lower than those for amorphous carbon (AC1 from *Rouleau and*

*Martin,* 1991), while β for organic-poor and silicate-rich (ROCK) particles exceed those for Mg-rich silicate with Fe inclusions. Thus, the values of β for ROCK particles and CHON particles are expected to be located somewhere between the values for amorphous carbon AC1 and those for Mg-rich silicate with Fe inclusions, although β for ROCK particles should be slightly lower than β for CHON particles. In the results, the spatial variations in the abundance of CHON and ROCK particles may not necessarily be noticeable in observations of cometary comae since β values of CHON particles do not seem to deviate from those of ROCK particles by a factor of 2.

In contrast, observational data may reveal the effect of porosity on the motion of dust particles in the comae, because β for highly porous and non-porous particles could differ by an order of magnitude or more if their size exceeds several micrometers. Therefore, if a variation in the optical properties of dust particles within a coma or among different comets is observed, the variation might be attributed to the difference in the porosities rather than the compositions.

Although β alone cannot constrain the physical properties of the dust particles, it can be used to check the validity of the characteristics of the cometary dust particles derived using other means. For example, an analysis of the temporal evolution in the ejecta of Comet 9P/Tempel 1 created by NASA's Deep Impact mission resulted in β ≈ 0.4, which suggests dust agglomerates of tens of micrometers in size consisting of silicate-core + organic-mantle monomers of 0.1-μm radius (*Kobayashi et al.,* 2013). It turned out that this interpretation is consistent with observational data on the color temperature, silicate feature strength, and degree of linear polarization (*Yamamoto et al.,* 2008). The values of β for dust particles in the tails of sungrazing comets derived from Solar and Heliospheric Observatory (SOHO) observations turned out to be 0.6 at maximum (*Sekanina,* 2000). This was found to be consistent with a predominance of Mg-rich pyroxene and olivine grains or small agglomerates of these grains in the tails of sungrazers remaining after sublimation of organic matter (*Kimura et al.,* 2002b). *Ishiguro et al.* (2016) derived the maximum value of β = 1.6 ± 0.2 from the optical observa-



tions of Comet 15P/Finlay and concluded that dust particles in the ejecta of multiple outbursts are fluffy agglomerates consisting of silicate-core and organic-mantle grains based on the numerical results of *Kimura et al.* (2003b).

## 4.2. Photoelectric Emission: Charging of Particles by Radiation

The surface of dust particles is inevitably charged, owing to the interactions of the particles with the UV and plasma. If free electrons are present on the surface of dust particles, then the boundary conditions of particle surface would be modified, which is the case for metallic particles. However, the surface of cometary dust is typically composed of dielectric materials such as ice, organic matter, and silicates. Because the electric conductivity of dielectric materials is extremely small compared to metals, the effect of surface charge on the scattering and absorption of light by cometary dust can be neglected. However, the surface charge might induce an observable effect on the plasma environments around comets through absorption and emission of electrons by the particles, which behave like super-heavy ions in a plasma.

When electromagnetic waves or photons are absorbed by a dust particle, photoelectrons might be emitted from the surface of the particle (*Feuerbacher et al.,* 1972). Photoelectron emission elevates the surface potential of the particle, predominating over other charging processes such as sticking of solar wind plasma and secondary electron emission (e.g., *Wyatt,* 1969; *Whipple,* 1981). The electric current due to photoelectron emission is a function of the absorption cross section and the photoelectric quantum yield, which is the number of photoelectrons per absorbed photon (e.g., *Mukai,* 1981; *Kimura and Mann,* 1998). Simple empirical formulae have been proposed to estimate the photoelectric quantum yield based on experimental data (e.g., *Draine,* 1978). However, the laboratory experiments tend to underestimate the true value of photoelectric quantum yields due to the geometrical limitations on the collection of photoelectrons in the laboratory (*Senshu et al.,* 2015).

Laboratory experiments on photoelectron emission are to a large extent limited to materials specific to the surfaces of satellites and spacecraft, but some measurements with astronomically relevant materials are available. *Baron et al.* (1978) conducted the experiments with a thin film of water ice at 13 and 80 K near the work function. As noted in the pioneering paper on the photoemission by *Einstein* (1905), the work function is a threshold, namely, the minimum energy required to move an electron to infinity (*Kittel,* 1995). *Feuerbacher et al.* (1972) measured the photoelectric quantum yield of glassy carbon, graphite, and lunar fines as a function of wavelength and determined the work function. The photoelectric quantum yields of submicrometer particles for silica, olivine, carbonaceous materials, and lunar samples were provided by *Abbas et al.* (2006, 2007), but the data might have suffered from calibration errors (*Kimura,* 2016).

The physical process of photoelectron emission is well

accounted for by a three-step model (*Smith,* 1971): (1) excitation of electrons by absorption of photons at a certain depth below the surface, (2) transportation of the excited electrons to the surface against inelastic collisions, and (3) ejection of the electrons, if their kinetic energies normal to the surface exceed the work function of the surface. A rigorous theory for estimating the photoelectric quantum yield requires the electric field inside the particle to be known. The photoelectric quantum yield has been so far computed only for spheres in the framework of Mie theory or geometrical optics (*Watson,* 1973; *Dwek and Smith,* 1996; *Kimura,* 2016). If dust particles are of submicrometer size or smaller, then the photoelectric quantum yield is in general enhanced due to the surface curvature of the particles and the contribution of the entire volume to absorption of light. On the contrary, the photoelectric quantum yield of very tiny nanometer-sized particles is reduced near the threshold energy of photons, owing to the enhancement of the work function for small particles (e.g., *Wong et al.,* 2003). The application of the DDA to the computation of the photoelectric quantum yield from agglomerates is straightforward, but no one has so far found worth a computation, which is most likely limited to the small particle sizes.

## 4.3. Radiative Torques: Grain Alignment and Rotational Disruption

The radiation-dust interaction can induce radiative torques (RATs) on a dust grain of non-spherical shape. It is found that RATs can cause the alignment of dust grains and their spinup. Scattering of light by aligned grains induces the circular polarization of scattered light, which may explain circular polarization in comets (see section 2.2.2). For a comprehensive reviews of those effects, see *Lazarian et al.* (2015) and *Andersson et al.* (2015). In cometary comae, RAT can spin up the dust particles to extremely fast rotation so they are disrupted by centrifugal stress, affecting the evolution of dust and ice in cometary coma (see *Hoang,* 2020, for a review).

The RAT induced by the radiation of wavelength λ acting on a grain of radius a is defined as

$$\Gamma_\lambda = \pi a^2 \gamma u_\lambda \left(\frac{\lambda}{2\pi}\right) \mathbf{Q}_\Gamma(\Theta, a, \lambda)$$

where γ is the anisotropy degree of the radiation field, $u_\lambda$ is the specific energy density, and $\mathbf{Q}_\Gamma(\Theta, a, \lambda)$ is the RAT efficiency vector that depends on the angle Θ between the radiation direction and the grain axis of maximum inertia moment (*Draine and Weingartner,* 1996; *Lazarian and Hoang,* 2007). The RAT efficiency vector has three components, $Q_{e1}$, $Q_{e2}$, $Q_{e3}$, with $\mathbf{e}_1$ chosen along the radiation direction $\mathbf{k}$.

### 4.3.1. Grain alignment and spinup by radiative torques.
The detailed study of grain alignment by RATs is presented in *Lazarian and Hoang* (2007). The authors found that RATs



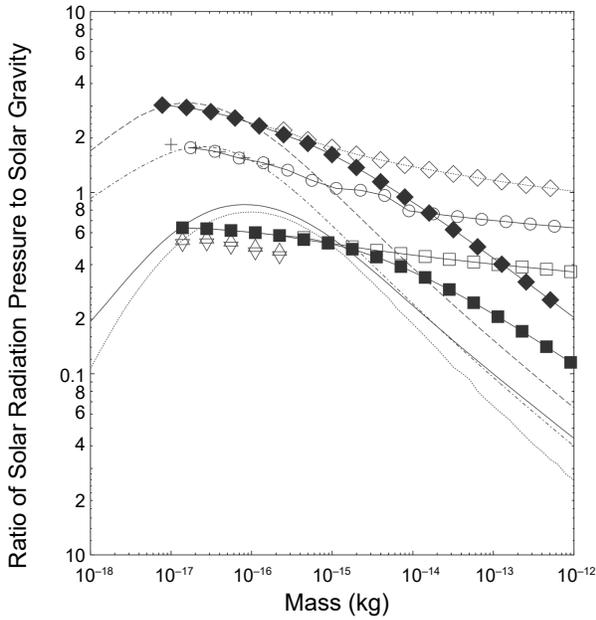

**Fig. 11.** The values of the parameter β for different dust particles: BPCAs (closed symbols) and BCCAs (open symbols) consisting of 0.1-μm-radius grains (see description of BPCA and BCCA in section 3.2.1). Different symbols represent amorphous carbon AC1 (diamonds), astronomical silicate (squares), Mg-rich olivine (triangles), Mg-rich pyroxene (inverse triangles), amorphous silicate with iron inclusions (circles), and silicate-core + organic-mantle (crosses). Lines show the results for solid spherical particles; the materials can be identified by the overlap of the line with the leftmost symbol.

can induce the alignment along the radiation direction (k-RAT) because the averaging over the fast rotation of the grain around the axis of major inertia $\mathbf{a}_1$ cancels out the align torque component $Q_{e2}$, leaving only the component $Q_{e1}$ that spins up the grain. When the grain initially makes an angle Θ with the radiation direction, the aligned component $Q_{e2}$ is not zero and can act to bring the grain back to the stationary point Θ = 0, establishing the k-RAT alignment (Fig. 12a).

If a magnetic field is present, then the axis of grain alignment depends on the precession of the grain angular momentum with the magnetic field and the radiation direction. If an external electric field is also present, under the effect of RATs, grain alignment can occur along one of these three axes, including the radiation direction, magnetic field, and electric field (Fig. 12b).

The magnetic properties control the efficiency of grain alignment in the magnetic field. Paramagnetic relaxation that was first suggested by *Davis and Greenstein* (1951) to align grains with the magnetic field is known as the Davis-Greenstein (D-G) mechanism. The paramagnetic relaxation is based on the dissipation of the rotational energy due to the rotating induced magnetization component perpendicular to the spinning axis, which ultimately leads to both minimum rotational energy with angular momentum and its magneti-

zation vector aligned with the ambient magnetic field only. The D-G mechanism implies that small grains are more efficiently aligned than large grains and that alignment becomes negligible when the grain rotational velocity is comparable to its thermal angular velocity (so-called thermal rotation). Furthermore, numerical simulations of D-G alignment for grains with different magnetic susceptibilities in *Hoang and Lazarian* (2016) demonstrate that, for superparamagnetic grains with embedded Fe inclusions, a maximum alignment degree of the grain axis with the magnetic field can be ~10–20%, although grains have superparamagnetic relaxation because the grain rotation velocity is smaller than the thermal value. This study establishes that efficient grain alignment is only achieved when the grain angular velocity exceeds its thermal angular velocity (i.e., suprathermal rotation), as previously suggested (e.g., *Purcell*, 1979). Such a requirement is satisfied by the spinup effect of RATs. Moreover, grains with enhanced magnetic susceptibility due to Fe inclusions can increase the paramagnetic relaxation rate that acts together with the RAT as well as increases the Larmor precession rate, which leads to B-RAT alignment (Fig. 12b).

**4.3.2. Effects of grain alignment on circular polarization in cometary comae.** As was shown in section 2.2.2, there are numerous observations of circular polarization of the scattered light in comets, and the most feasible mechanism to describe it is alignment of the dust particles by radiative torque in the presence of the solar magnetic field. This mechanism is considered in this section.

The intensity of circularly polarized radiation resulting from scattering on a dust particle is given by the fourth Stokes parameter V (see section 2.2) and is equal to V = $I_0 S_{41}/k^2 R^2$ where $S_{41}$ is the 4–1 element of the Mueller matrix, $I_0$ is the intensity of the incident light, k is the wave number, and R is the distance to the source of light (e.g., heliocentric distance). Using DDA and considering the grains with x < 1 (Rayleigh particles), *Hoang and Lazarian* (2014) showed that

$$V = \frac{I_0 k^4}{2R^2} i(\alpha_\| \alpha^*_\perp - \alpha^*_\| \alpha_\perp)([\mathbf{e}_{inc} \times \mathbf{e}_{sca}].\mathbf{e})(\mathbf{e}_{inc}.\mathbf{e})$$

where $\mathbf{e}_{inc}$ and $\mathbf{e}_{sca}$ are the unit vectors of incident (**k**) and scattering direction, $\alpha_\|$ and $\alpha_\perp$ are the complex polarizabilities along the grain alignment axis **e** and in the perpendicular direction, respectively. From the above equation one can see that circular polarization can be produced only when the grain has an absorbing material, i.e., its polarizability (or refractive index) contains a non-zero imaginary part. Moreover, circular polarization is not equal to zero only when the grain alignment axis **e** is not parallel to the incident direction $\mathbf{e}_{inc}$. If **e** ∥ $\mathbf{e}_{inc}$, the cross-vector [$\mathbf{e}_{inc} \times \mathbf{e}_{sca}$] becomes perpendicular to **e**, and since V ∝ [$\mathbf{e}_{inc} \times \mathbf{e}_{sca}$] · **e**, in this case V = 0.

Thus, the k-RAT alignment itself produces zero circular polarization as it aligns the dust particles in the direction where the alignment axis **e** is along the incoming radiation $\mathbf{e}_{inc}$.

However, the presence of a magnetic field induces a magnetic torque on the grain magnetic dipole moment,

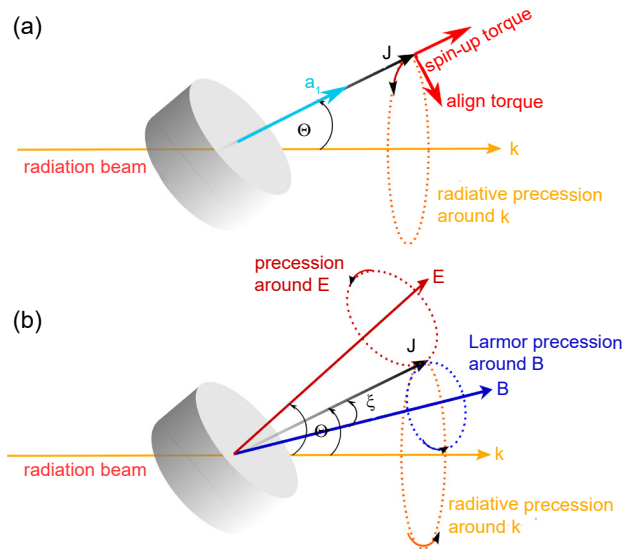

**Fig. 12.** Illustration of grain alignment and spin-up by radiative torques assuming that the grain shortest axis $a_1$ is aligned along the angular momentum **J**. **(a)** The grain angular momentum **J** experiences fast radiative precession along the radiation beam **k**. The spinup torque component along the angular momentum acts to spin the grain up, while the align torque component acts to bring **J** in alignment with **k**, resulting in the so-called k-RAT alignment. **(b)** In the presence of the magnetic field (**B**) and electric field (**E**), the grain experiences the precession around **B** (Larmor precession), around the electric field, and the radiation direction, if the grain has a component of the magnetic moment and electric dipole moment parallel to **J**.

which causes Larmor precession of the grain angular momentum around the magnetic field (Fig. 12b). If the Larmor precession is faster than the radiative precession, the magnetic field becomes the axis of grain alignment (B-RAT), and the grain short axis **e** deviates from the radiation direction $e_0$, producing non-zero circular polarization. To have the precession caused by the magnetic field efficient in providing a deviation from the radiative direction, it should be faster than the precession caused by RAT. For solar radiation at the heliocentric distance ~1 au, *Hoang and Lazarian* (2014) showed that the RAT alignment time is $t_{RAT} = 3 \times 10^4$ (a/0.1 μm)$^{1/2}$ s, i.e., for particles of radius a = 0.1 μm, $t_{RAT} = 3 \times 10^4$ s and for particles of radius a = 10 μm, $t_{RAT} = 3 \times 10^5$ s. Thus, although small paramagnetic particles can be efficiently aligned by RAT combined with the magnetic field, alignment of larger particles requires larger values of the magnetic susceptibility. This can be provided by the presence of superparamagnetic, ferromagnetic, and ferrimagnetic inclusions, which increase grain Larmor precession rate and make this mechanism more efficient (*Hoang and Lazarian,* 2016). For a dust particle containing clusters of metallic atoms with $N_{cl} < 10^5$ atoms per cluster, the magnetic susceptibility increases by a factor of $N_{cl}$, decreasing the timescales of magnetic relaxation and Larmor precession by the same factor. As a result, grain alignment with the magnetic field (B-RAT) is more likely than the k-RAT alignment. *Hoang and Lazarian* (2016) have demonstrated that even a small fraction of metal present in the form of metallic nanoparticles, for example, GEMS present in cometary dust (e.g., *Keller and Messenger,* 2011), can provide the perfect alignment of dust grains that are subject to RATs. The considered mechanism was used to explain circular polarization in comets by *Kolokolova et al.* (2016), who showed that the penetration of the solar magnetic field into cometary coma can provide the alignment of dust particles that causes circular polarization of the light scattered by cometary dust.

**4.3.3. Rotational disruption by radiative torques.** RATs are very efficient in spinning up a non-spherical grain to suprathermal rotation, i.e., rotation with rates above the thermal rotation rate (*Draine and Weingartner,* 1996; *Hoang and Lazarian,* 2009). *Hoang et al.* (2019) realized that centrifugal stress resulting from such suprathermal rotation can exceed the maximum tensile strength of grain material ($S_{max}$), resulting in the disruption of the grain into fragments. This new physical mechanism was termed radiative torque disruption (RATD). Since rotational disruption acts to break the loose bonds between the monomers, unlike breaking the strong chemical bonds between atoms in the case of thermal sublimation, RATD can work for the solar radiation field beyond the sublimation zone of refractory dust (*Hoang,* 2020). For the cometary coma, *Hoang* (2021) showed that the disruption size, i.e., the minimum radius starting from which the particles can be disrupted, is $a_{disr} \sim 0.6$ μm for the gas density of $n_H = 10^{10}$ cm$^{-3}$ and decreases with increasing cometocentric radius or decreasing gas density. There exists a maximum radius of grains that can still be disrupted by centrifugal stress (*Hoang and Tram,* 2020) of $a_{disr,max} \sim 2.2$ μm for the gas density of $n_H = 10^{10}$ cm$^{-3}$, assuming $S_{max} = 10^7$ erg cm$^{-3}$.

*Hoang and Tung* (2020) showed that RATD is efficient in disrupting large composite grains into smaller ones. Therefore, the RATD mechanism implies the evolution of dust properties, causing a decrease in the abundance of large grains and increase in the abundance of small ones with the cometocentric and heliocentric distances. The efficiency of RATD also increases with decreasing the gas production rate $Q_{gas}$ because the latter determines the rotational damping of grains spun up by RATs. *Herranen* (2020) studied the rotational disruption of agglomerates of radius ~2 μm in sungrazing comets and found that the mechanism could be efficient for those grains. Thus, the size and structure change in the cometary dust under RATD may explain the polarization properties reported in *Thompson* (2020) for Comet ISON.

One more effect of RATs on cometary dust is the rotational desorption of icy mantles on the particles. The current model of ice removal from the dust grains is based on thermal sublimation from icy grains (*Cowan and A'Hearn,* 1979). However, *Hoang and Tung* (2020) found that water ice mantles could be desorbed from the grain core by radiative torques. The rotational desorption can occur at heliocentric distances of $R_{des} \sim 20$ au, much larger than the water sublimation radius, $R_{sub}(H_2O) \sim 3$ au, thus making the RATD mechanism more efficient than sublimation at large heliocentric distances.

## 5. FINAL REMARKS AND FUTURE WORK

Our understanding of the properties of cometary dust has changed dramatically since 2004 when *Comets II* was published. The chapter in that volume by *Kolokolova et al.* (2004a) ended with the conclusion of a high likelihood of the aggregated structure of the cometary dust particles. Now this is a well-established fact, mainly proven by Stardust and Rosetta mission studies of cometary dust but also supported by computer and laboratory simulations, which have shown that cometary observations, especially polarimetric ones, can be most successfully reproduced if the dust is modeled as a mixture of agglomerated particles of different porosity, from very fluffy to rather compact, which is consistent with the mission-result characteristics of the cometary dust summarized in *Güttler et al.* (2019).

Modeling of agglomerates is in high demand in various studies of cometary dust and nuclei, e.g., in computing radiation pressure, radiative torque, tensile strength, and thermal properties of cometary material. This requires paying due attention to the computer modeling of the realistic cometary dust particles.

One of the main characteristics of a model should be its capability to simulate a comprehensive set of observational data. Polarization phase curve, brightness phase function, and color data in isolation are all ambiguous, and a model that reproduces only one of them has a high probability of being wrong. Thus, modelers always need to check the consistency of their models with all available observational data, angular and spectral, including those in the NIR and thermal infrared as well as with the dynamical properties of the dust particles derived from the morphology of the coma. Also, models should be consistent with ideas about the formation of dust particles in the protosolar nebula and with the cosmic abundances of the elements. Finally, if the images of the coma, including polarimetric maps, are available, the model should be checked for the capability of reproducing the observed trends using physically plausible assumptions on particle fragmentation and sublimation of its material.

We also would like to point out that by using a very complex model with numerous parameters, one can fit almost any observational data. If a model varies the size, structure, and composition of the particles and considers several populations of particles in the dust, a good fit to the observations can be found, but it may not present the realistic characteristics of the cometary dust. To avoid this trap, it is highly recommended to limit the number of model parameters using, where possible, information achieved by cometary missions. The Stardust and Rosetta missions significantly extended our knowledge of cometary dust particles, and the chapter in this volume by Engrand et al. presents a detailed review of those findings. A helpful source of information can be also a comprehensive review of the *in situ* and laboratory data by *Levasseur-Regourd et al.* (2018).

We expect further progress in the computer modeling of the interaction of cometary particles with radiation. The computational tools, specifically such codes as DDA, MSTM, FaSTMM, $R^2T^2$, etc., are sufficiently powerful to model cometary dust. However, they need more computational resources to be capable of modeling realistically large (hundreds of micrometers) and complex (e.g., hierarchical) particles, which will come as more powerful computers and computer clusters become available.

A better understanding of the interaction of cometary dust and surfaces with radiation will be produced by laboratory modeling, specifically models at low temperatures (see the chapter in this volume by Poch et al.), and by the new theoretical and laboratory developments regarding comet dust formation and evolution (see the chapters in this volume by Aikawa et al. and Simon et al.).

Remote sensing of cometary dust and surfaces remains one of the major resources of information provided by space missions using cameras working in different spectral ranges. For example, the future mission Comet Interceptor (see the chapter in this volume by Snodgrass et al.) has planned several instruments for remote sensing of the cometary environment, among them the Optical Periscopic Imager for Comets (OPIC) and Entire Visible Sky Camera (EnVisS), the latter of which has polarimetric capabilities, along with the Modular Infrared Molecules and Ices Sensor (MIRMIS) for the wavelength range 0.9–25 μm, which, together with the other cameras, may provide significant constraints on dust models. Also, observational information on more comets is expected, allowing statistical analysis of the dust properties and comet classification based on the properties of its dust that may come from the observations with new telescopes such as the James Webb Space Telescope (JWST) as well as groundbased and spacebased surveys (see the chapter in this volume by Bauer et al.). Studies that span a wide range of wavelengths and simultaneously use different techniques (e.g., thermal emission and scattered light) will help to understand the observed comet-to-comet distribution of cometary dust properties.

Finally, important steps forward in remote sensing of comets will be development in two main areas: (1) broadening the wavelength range of observations, which should include more UV observations (sensitive to nanoparticles if present) and microwave radiation capable of providing information about large, millimeter-sized, dust particles; and (2) expanding the observations to other objects related to the solar system comets such as debris disks and interstellar and exosolar comets.

***Acknowledgments.*** L.K. acknowledges helpful discussions with J. Markkanen, N. Kiselev, and O. Shubina. This work was partially supported by NASA Grants No. 80NSSC21K0164 and 80NSSC17K0731 and Grants-in-Aid for Scientific Research (KA-KENHI #21H00050) of JSPS.